\documentclass{aa}  

\usepackage{graphicx}
\usepackage{subcaption}
\usepackage{txfonts}
\def\PLUTO{{\sc pluto}}
\def\FLASH{{\sc flash}}

\newcommand{\sna}{SN 1987A}

\begin{document} 

  \title{Tracing the ejecta structure of SN\,1987A: \\ Insights and diagnostics from 3D MHD simulations}




      \author{S.\,Orlando\inst{1}
      \and M.\,Miceli\inst{2,1}
      \and M.\,Ono\inst{3,4}
      \and S.\,Nagataki\inst{4,5,6}
      \and M.-A.\,Aloy\inst{7,8}
      \and F.\,Bocchino\inst{1}
      \and M.\,Gabler\inst{7}
      \and B.\,Giudici\inst{7}
      \and R.\,Giuffrida\inst{9,1}            
      \and E.\,Greco\inst{1}
      \and G.\,La Malfa\inst{7,1}
      \and S.-H.\,Lee\inst{10}
      \and M.\,Obergaulinger\inst{7}
      \and O.\,Petruk\inst{1,11}
      \and V.\,Sapienza\inst{1}
      \and \\S.\,Ustamujic\inst{1}
      \and J. Weng\inst{12,1}
}

    \institute{INAF -- Osservatorio Astronomico di Palermo, Piazza del Parlamento 1, I-90134 Palermo, Italy\\
    \email{salvatore.orlando@inaf.it}
    \and Dip. di Fisica e Chimica, Universit\`a degli Studi di Palermo, Piazza del Parlamento 1, 90134 Palermo, Italy
    \and Institute of Astronomy and Astrophysics, Academia Sinica, No.1, Sec.4, Roosevelt Road, Taipei 106319, Taiwan
    \and Astrophysical Big Bang Laboratory (ABBL), RIKEN Pioneering Research Institute (PRI), 2-1 Hirosawa, Wako, Saitama 351-0198, Japan
    \and RIKEN Center for Interdisciplinary Theoretical \& Mathematical Sciences (iTHEMS), 2-1 Hirosawa, Wako, Saitama 351-0198, Japan
    \and Astrophysical Big Bang Group (ABBG), Okinawa Institute of Science and Technology (OIST), 1919-1 Tancha, Onna-son, Kunigami-gun, Okinawa 904-0495, Japan
    \and Departament d’Astonomia i Astrof\'isca, Universitat de Vale\`encia, E-46100 Burjassot (Val\`encia), Spain
    \and Observatori Astron\`omic, Universitat de Val\`encia, 46980 Paterna, Spain
    \and Universit\'e Paris-Saclay, Universit\'e Paris Cit\'e, CEA, CNRS, AIM, 91191 Gif-sur-Yvette, France
    \and{Department of Astronomy, Kyoto University, Oiwake-cho, Kitashirakawa, Sakyo-ku, Kyoto 606-8502, Japan}
    \and{Institute for Applied Problems in Mechanics and Mathematics, Naukova Str. 3-b, 79060 Lviv, Ukraine}
    \and{School of Astronomy and Space Science, Nanjing University, 163 Xianlin Avenue, Nanjing 210023, China}
             }

   \date{Received September 15, 1996; accepted March 16, 1997}

 
  \abstract
   {Supernova (SN) 1987A provides a unique window into the aftermath of a massive stellar explosion, offering key insights into the ejecta's morphology, composition, explosion mechanism, progenitor system, and circumstellar medium (CSM) interaction.}
   {This study employs high-resolution three-dimensional magnetohydrodynamic (3D MHD) simulations to investigate large-scale ejecta asymmetries in \sna. By comparing the simulations with JWST observations and making predictions for XRISM, we aim to refine our understanding of the explosion mechanism and the remnant’s evolution.}
   {We performed 3D MHD simulations that trace the evolution of \sna\ from the SN to the SNR, extending our predictions up to 5000 years into the future and considering the Ni-bubble effects. The simulation results are compared with JWST observations and used to predict XRISM spectra, providing a means to evaluate the accuracy of the modeled ejecta structure.}
   {Our simulations reproduce the large-scale Fe-rich ejecta morphology observed with JWST, revealing two prominent clumps suggestive of a bipolar explosion. The Ni-bubble effect in the first year enhances Fe-rich ejecta expansion, accelerating their interaction with the reverse shock. However, discrepancies with JWST observations in clump velocities and spatial distribution suggest stronger explosion asymmetries than modeled. Since 2021, our models predict that shocked ejecta have contributed increasingly to X-ray emission, now rivaling shocked CSM and soon dominating as the latter fades. Future XRISM observations will trace the evolution of these ejecta structures, refining constraints on explosion geometry. Early remnant asymmetries from CSM interaction may persist for at least 100 years.}
   {Our findings reinforce the role of highly asymmetric core-collapse mechanisms in shaping \sna’s ejecta and provide critical constraints on explosion geometry. Future studies should investigate more extreme explosion asymmetries, potentially arising from stochastic processes in neutrino-driven core collapse or magneto-rotational SN models, to identify the mechanism that best explains \sna’s nearly bipolar Fe-rich ejecta structure.}


\titlerunning{Tracing the ejecta structure of SN\,1987A through 3D MHD Simulations}
\authorrunning{S. Orlando et~al.}

   \maketitle
%

\section{Introduction}

Supernova (SN) 1987A is a unique laboratory, providing an unparalleled opportunity to study the connection between a massive progenitor star, its terminal SN explosion, and the subsequent evolution of the supernova remnant (SNR) interacting with the surrounding circumstellar medium (CSM). Located in the Large Magellanic Cloud at a distance of $\approx 51.4$~kpc \citep{1999IAUS..190..549P} and monitored for nearly four decades \citep[e.g.,][ and references therein]{2021ApJ...916...41S, 2024ApJ...966..147R}, \sna\ offers critical insights into the physics of core-collapse SNe, the mechanisms shaping their remnants, and the imprints of the progenitor star’s properties on these processes.

A distinct characteristic of \sna\ is its highly asymmetric ejecta structure \citep[e.g.,][]{2013ApJ...768...89L, 2016ApJ...833..147L}, along with its peculiar ring-like morphology observed across multiple wavelengths \citep[e.g.,][]{2010ApJ...722..425D, 2011Natur.474..484L, 2012A&A...548L...3M, 2013ApJ...764...11H}. This morphology is widely attributed to the interaction of the expanding remnant with an inhomogeneous CSM \citep[e.g.,][]{1995ApJ...452L..45C}, making it a valuable probe of the progenitor’s mass-loss history. The presence of a characteristic triple-ring nebula \citep{1989ApJ...347L..61C} and density variations in the surrounding medium suggest episodic, non-uniform mass loss, potentially linked to a binary merger scenario \citep[e.g.,][]{2007Sci...315.1103M}.

The unshocked ejecta, which have yet to encounter the reverse shock, provide direct insight into the explosion mechanism and progenitor structure. Observations of iron \citep[Fe; e.g.,][]{1990ApJ...360..257H, 2023ApJ...949L..27L} and titanium  \citep[Ti; e.g.,][]{2015Sci...348..670B} lines, as well as the elliptical distribution of the inner ejecta \citep{2013ApJ...768...89L, 2016ApJ...833..147L} and molecular tracers such as CO and SiO \citep{2017ApJ...842L..24A}, indicate a highly anisotropic explosion, likely shaped by instabilities during core collapse. These features encode key information about explosion dynamics, including the standing accretion shock instability (SASI) and neutrino-driven convection \citep[e.g.,][]{2003ApJ...584..971B, 2017hsn..book.1095J, 2021Natur.589...29B, 2025A&A...696A.108O}. Conversely, shocked ejecta offer crucial insights into the remnant's interaction with the inhomogeneous CSM and the structure of the outermost ejecta impacted by the reverse shock. This, in turn, provides key information on the progenitor star’s mass-loss history and its nature (e.g., \citealt{2022A&A...666A...2O, 2025A&A...696A.188O}).

Advances in observational capabilities, particularly with the James Webb Space Telescope (JWST), are providing unprecedented insights into \sna’s ejecta structure and composition. Recent findings reveal a highly asymmetric [Fe\,I] distribution, resembling a broken dipole, with clumps moving at $\sim 2300$~km~s$^{-1}$ and Fe-rich ejecta interacting with the reverse shock \citep{2023ApJ...949L..27L}. The reverse shock, traced by He\,I emission, forms a bubble-like structure with a $\sim 45^{\circ}$ half-opening angle. NIRCam imaging has identified substructures within the ejecta ("the bar"), faint H$_{2}$ crescents between the ejecta and the equatorial ring, and bright $3-5\,\mu$m continuum emission beyond the ring, driven by synchrotron radiation and dust \citep{2024MNRAS.532.3625M}. Additionally, JWST spectroscopy has detected blueshifted, spatially unresolved infrared emission lines of argon (Ar) and sulfur (S) from the innermost ejecta, likely ionized by a cooling neutron star or pulsar wind nebula, with a velocity shift suggestive of a potential neutron star natal kick \citep{2024Sci...383..898F}. Complementary observations with XRISM in the X-ray band have started to further constrain high-energy emission of \sna\ and the interactions between the CSM and shocked ejecta \citep[e.g.,][]{2024ApJ...961L...9S}.

Interpreting these unprecedented high-quality data requires advanced modeling tools to extract key insights into the parent SN and progenitor star. Neutrino-driven SN explosions generally produce multi-polar asymmetries, but some may result in unipolar or strongly bipolar morphologies \citep[e.g.,][]{2015A&A...577A..48W, 2021ApJ...915...28B, 2024ApJ...974...39W, 2024ApJ...964L..16B}, potentially explaining the features observed in \sna, including the neutron star’s kick direction (\citealt{2017ApJ...837...84J, 2017IAUS..331..148J}). In fact, the development of extended Ni-rich plumes can be influenced by progenitor characteristics, explosion energy, and stochastic processes, resulting in a wide range of ejecta structures. Other theoretical models propose bipolar or jet-like SN explosions driven by alternative mechanisms, including strong magnetic fields and rotation \citep[e.g.,][]{2002ApJ...568..807W, 2006A&A...450.1107O, 2009A&A...498..241O, 2009ApJ...691.1360T, 2014ApJ...785L..29M, 2018MNRAS.478..682B, 2024Galax..12...29S, 2024RAA....24g5006S}. Some studies suggest that asymmetric, bipolar-like explosions can efficiently produce Ti at levels of $\sim 10^{-4}\,M_{\odot}$ \citep{1997ApJ...486.1026N} and account for the high-velocity Fe component ($\sim 3000$~km~s$^{-1}$) observed in \sna\ \citep{1994ApJ...427..874C, 1998ApJ...495..413N}. On the other hand, comparable Ti yields and Fe ejecta velocities can be also obtained in simulations of neutrino-driven SN explosions (e.g., \citealt{2015A&A...581A..40U, 2019A&A...624A.116U, 2021ApJ...914....4U, 2023ApJ...957L..25S, 2024ApJ...974...39W}). Additionally, the asymmetric Fe line profile in this remnant \citep{1990ApJ...360..257H} has been linked to a bipolar explosion with asymmetry relative to the equatorial plane, offering a possible explanation for the neutron star’s kick direction \citep{2000ApJS..127..141N}.

In more recent years, 3D magnetohydrodynamic (MHD) simulations have proven to be powerful tools for extracting information about the SN explosion and progenitor, playing a crucial role in interpreting multi-wavelength observations. These models have successfully reproduced key characteristics of \sna, including: (i) the explosion’s asymmetries and energetics (\citealt{2020ApJ...888..111O, 2020MNRAS.494.2471J}), as well as an estimate of the neutron star’s kick velocity, with the latter in good agreement with recent observations of the point source discovered in \sna\ (\citealt{2024Sci...383..898F}); (ii) the signatures of the progenitor star on the remnant structure at the age of tens of years (\citealt{2020A&A...636A..22O}, hereafter Paper I); (iii) the CSM’s structure (\citealt{2015ApJ...810..168O} and Paper I) and pre-SN magnetic field configuration (\citealt{2019A&A...622A..73O, 2023MNRAS.518.6377P}); and (iv) non-thermal emission from a putative pulsar wind nebula embedded in cold, dense ejecta near the remnant’s center \citep{2021ApJ...908L..45G, 2022ApJ...931..132G}. An alternative scenario to the last point has been proposed by theoretical studies, suggesting that the infrared excess observed near the predicted position of the compact object in \sna\ may be most plausibly attributed to thermal emission from a young, cooling neutron star \citep{2020ApJ...898..125P, 2023ApJ...949...97D}. The 3D MHD simulations also predicted the current transition to a new evolutionary phase dominated by X-ray emission from shocked ejecta, a prediction recently confirmed by Chandra \citep{2024ApJ...966..147R} and XMM-Newton \citep{2025ApJ...981...26S}.

In this work, we expand on previous studies by investigating the complex ejecta structure of \sna\ through 3D MHD simulations that reproduce most of its observable features. The inclusion of magnetic fields is crucial, as they play a key role in stabilizing the dense equatorial ring against complete disruption by hydrodynamic instabilities that develop at the ring's boundary following the passage of the forward shock (\citealt{2019A&A...622A..73O}). Our primary objective is to compare these models with recent JWST observations, identifying both consistencies and discrepancies that could highlight missing physical processes or provide new constraints on the progenitor SN and CSM properties. This analysis can be crucial for refining our understanding of the SN explosion mechanism and remnant evolution. We also provide detailed predictions for XRISM observations, focusing on the detectability of shocked ejecta and exploring ejecta diagnostics that will soon be testable with XRISM’s high spectral-resolution data. Finally, we investigate the future evolution of \sna, examining whether signatures of its early interaction with a highly inhomogeneous CSM or evidence of an asymmetric explosion may still be observable in older remnants with similar characteristics.

The paper is organized as follows: Sect.~\ref{sec:model} details the numerical setup of our 3D MHD simulations; Sect.~\ref{sec:result} presents simulation results and their comparison with observations; Sect.~\ref{sec:summary} discusses the implications of our findings and summarizes key conclusions.

\section{The 3D MHD model}
\label{sec:model}

The 3D MHD simulations of \sna, which trace its evolution from the initial explosion to the formation of the SNR over 50 years, were extensively presented in Paper I. These simulations integrate 3D SN models (\citealt{2020ApJ...888..111O}) that capture the asymmetric explosion from core collapse to shock breakout at the stellar surface, with 3D MHD SNR simulations (e.g., \citealt{2019A&A...622A..73O}) that follow the subsequent evolution and interaction with an inhomogeneous CSM. Key physical processes, including the mixing and clumping of chemically homogeneous ejecta layers, their interaction with the structured CSM, and the transition from the SN to the SNR phase, were carefully modeled. Synthetic observables enabled direct comparisons with observations, constraining the explosion dynamics, CSM structure, and progenitor properties

In this paper, we extended the simulations up to an age of 5000 years, providing predictions for the long-term evolution of \sna. Rather than serving as a direct template for future observations of \sna\ (which will be conducted with much more different observational capabilities), this analysis aims to explore features that may be visible today in other, more evolved remnants shaped by a CSM similar to that of \sna. We also investigated the impact of the "Ni-bubble effect" (\citealt{1992ApJ...387..294H}; which was not considered in our previous \sna\ simulations), driven by the radioactive decay of nickel (Ni) synthesized during the explosion (see \citealt{2021A&A...645A..66O}), on the remnant's dynamics. Below, we summarize the key components and physical processes incorporated into our model. For further details on the underlying physics and numerical implementation, we refer the reader to Paper I.

\subsection{Pre-supernova models}
\label{sec:stellar_mod}

The progenitor of \sna, Sanduleak (Sk) $-69^{\circ}202$, was a blue supergiant (BSG) with an initial mass of approximately $20\,M_{\odot}$. In Paper I, we investigated three pre-SN models to explore the evolution of \sna: (i) a $16.3\,M_{\odot}$ BSG evolved as a single star with a compact hydrogen (H) envelope and a helium (He) core mass consistent with Sk $-69^{\circ}202$ (model N16.3; \citealt{1988PhR...163...13N}); (ii) a $18.3\,M_{\odot}$ BSG resulting from the merger of two massive stars with masses of $14\,M_{\odot}$ and $9\,M_{\odot}$, respectively (model B18.3; \citealt{2018MNRAS.473L.101U}); (iii) a $19.8\,M_{\odot}$ red supergiant (RSG) with an extended H envelope (model S19.8; \citealt{2018ApJ...860...93S}), representing a progenitor with a structure in the He shell and H envelope significantly different from that of a BSG. This latter model was considered to test whether the imprint of the progenitor star can be identified in the structure of the SNR.

The analysis presented in \cite{2020ApJ...888..111O} and Paper I demonstrated that the merger model B18.3 best reproduces the observed properties of the remnant of \sna. Specifically, the remnant originating from this model most accurately matches the broadening and Doppler shifts of Fe (\citealt{1990ApJ...360..257H}) and Ti (\citealt{2015Sci...348..670B}) lines observed in the years following the SN explosion. These spectral features provide critical insights into the dynamics of the ejecta and the asymmetries introduced during the explosion (\citealt{2020MNRAS.494.2471J}). Based on these findings, we focused exclusively on the B18.3 model in the present study.

\subsection{Modeling the supernova evolution}

The SN phase was modeled using the \FLASH\ hydrodynamic code \citep{for00} on a 3D Cartesian grid (see \citealt{2020ApJ...888..111O} for details). The simulations employed the pre-SN models described in Sect.~\ref{sec:stellar_mod} at the time of core-collapse as initial conditions and considered key physical processes, including gravity (both self-gravity and the gravitational influence of the proto-neutron star), fallback of material onto the proto-neutron star, and a nuclear reaction network with 19 species (neutrons, protons, $^{1}$H, $^{3}$He, $^{4}$He, $^{12}$C, $^{14}$N, $^{16}$O, $^{20}$Ne, $^{24}$Mg, $^{28}$Si, $^{32}$S, $^{36}$Ar, $^{40}$Ca, $^{44}$Ti, $^{48}$Cr, $^{52}$Fe, $^{54}$Fe, and $^{56}$Ni) to track nucleosynthesis. The model also accounted for the feedback of nuclear energy generation and the energy deposition from the radioactive decay of $^{56}$Ni synthesized in the explosion. The equation of state was tailored to different physical regimes: the Helmholtz equation of state \citep{2000ApJS..126..501T} was applied in early phases characterized by high densities and temperatures, while a radiation and ideal gas equation of state was used for later phases \citep{2013ApJ...773..161O}.

The explosion was initiated by injecting thermal and kinetic energy near the interface between the Fe core and silicon (Si) layer. Asymmetries were introduced to emulate anisotropic mechanisms such as the standing accretion shock instability (SASI; \citealt{2003ApJ...584..971B}), magnetic jet-driven explosions \citep{1999ApJ...524L.107K, 2020MNRAS.492.4613O, 2021MNRAS.503.4942O}, or single-lobe SN geometries \citep{2005ApJ...635..487H}. Different asymmetry configurations, explored in \cite{2020ApJ...888..111O}, were implemented by modulating the initial radial velocity distribution and introducing perturbations. The study demonstrated that a bipolar-like explosion with asymmetry across the equatorial plane (resulting in greater energy release on one side) and a total energy of $2\times 10^{51}$~erg (see Table~1 in Paper I) best reproduced the observed shifts and broadening of Fe lines 1–2 years post-core collapse, aligning with the elliptical morphology of the inner ejecta distribution in \sna\ \citep{2013ApJ...768...89L, 2016ApJ...833..147L, 2023ApJ...949L..27L}. 

To resolve the large spatial scales associated with the remnant expansion, a mesh strategy analogous to that employed by \citet{2013ApJ...773..161O} was adopted. The initial computational domain spanned $-5000$ to $5000$~km in the $x$, $y$, and $z$ directions, centered on the SN origin, and encompassed the oxygen-rich (O) layer of the progenitor star. The simulations employed adaptive mesh refinement (AMR; \citealt{2000CoPhC.126..330M}) to achieve a maximum resolution of $\sim 10$~km. As the forward shock approached the domain boundaries, the computational domain was dynamically extended by a factor of 1.2 in all directions, with physical quantities re-mapped onto the new domain and the extended regions initialized to pre-SN conditions. This strategy required approximately 75 re-mappings to track the shock wave propagation through the star and its breakout at the stellar surface, occurring about 20 hours after core collapse. The final domain extended from $-2\times 10^{14}$ to $2\times 10^{14}$~cm in all directions, with a finest spatial resolution of $2\times 10^{11}$~cm.

\subsection{Modeling the transition to the SNR phase}
\label{sec:snrmod}

The output of the SN simulations, approximately 20 hours after core collapse, was used as the initial condition for 3D MHD simulations of the blast wave interacting with the ambient medium. The numerical setup is described in Paper I. The time-dependent MHD equations for mass, momentum, energy, and magnetic flux conservation are solved using the \PLUTO\ code \citep{2007ApJS..170..228M, 2012ApJS..198....7M}. The simulations account for deviations from electron-ion temperature equilibration, departures from ionization equilibrium \citep{2015ApJ...810..168O}, and the chemical evolution of the ejecta using a multi-fluid approach \citep{2016ApJ...822...22O}. Each fluid tracked a specific species from the SN simulations, while the adopted CSM abundances are those derived from the analysis of X-ray observations \citep{2009ApJ...692.1190Z}. 

In addition to the simulations presented in Paper I, a new model (B18.3-Dec; see Table~\ref{Tab:model}) was introduced, based on the setup of model B18.3 presented in Paper I but including energy deposition from the radioactive decay chain $^{56}$Ni $\rightarrow$ $^{56}$Co $\rightarrow$ $^{56}$Fe. The decay energy was treated as a local energy source, neglecting neutrino contributions and assuming no $\gamma$-ray leakage from the inner part of the remnant (see \citealt{2021A&A...645A..66O} for the details of the implementation). 

The adopted CSM model reproduces the highly inhomogeneous triple-ring nebula with its characteristic hourglass shape, as revealed by multiwavelength observations (e.g., \citealt{2005ApJS..159...60S}). Specifically, in the immediate surrounding of the progenitor star ($r < 0.08$~pc), the CSM was modeled as a spherically symmetric stellar wind with a density profile $\propto r^{-2}$, a mass-loss rate $\dot{M}_{\rm w} = 10^{-7} M_{\odot}$~yr$^{-1}$, and a velocity $u_{\rm w} = 500$~km~s$^{-1}$ \citep{2007Sci...315.1103M}. Although the CSM structure in the vicinity of the progenitor may, in reality, be more complex, we assumed spherical symmetry. This simplification is justified, as any asymmetries in the tenuous wind at these small scales would have negligible effects on the blast wave dynamics, which remains in the free-expansion phase. At a distance of $r_{\rm HII} = 0.08$~pc in the equatorial plane, the wind interacts with a nebula consisting of an H\,II region, a dense equatorial ring, and two less dense polar (outer) rings (\citealt{1995ApJ...452L..45C, 2005ApJS..159...60S}; see also Fig.~1 in \citealt{2015ApJ...810..168O}). This CSM component plays a crucial role in the subsequent evolution of the remnant and in shaping its morphology. The equatorial ring contained both a smooth component and high-density clumps, whose parameters were explored around fiducial values from \citet{2015ApJ...810..168O}. Following \cite{2005ApJS..159...60S}, the polar rings are described as inward-pointing spurs with a density comparable to that of the H\,II region, located $1.2\times 10^{18}$~cm above and below the equatorial plane. This places the innermost regions of the outer rings approximately $1.7\times 10^{18}$~cm from the explosion center, consistent with optical observations (\citealt{2011A&A...527A..35T}). The density of the H\,II region declines as $r^{-2}$ for distances greater than $\approx 2.2\times 10^{18}$~cm. The parameters of the CSM adopted for models B18.3 and B18.3-Dec are the same and are summarized in Table~2 in Paper I. Passive tracers were used to track the evolution of ejecta, H\,II region, and ring material, as well as shock properties required for synthesizing multi-wavelength emission.

\begin{table}
\caption{Setup for the simulated models.}
\label{Tab:model}
\begin{center}
\begin{tabular}{llll}
\hline
\hline
Model       & Progenitor & $E_{\rm exp}$   & rad.     \\ 
            & system     & [$10^{51}$ erg] & decay    \\ \hline
B18.3$^{a}$ & merger & 2.0   & no  \\
B18.3-Dec   & merger & 2.0   & yes \\
\hline
\end{tabular}
\end{center}
$(a)$ Model presented in Paper I.
\end{table}

As demonstrated by \cite{2019A&A...622A..73O}, the magnetic field plays a crucial role in the evolution of \sna, particularly in the dynamics of the dense equatorial ring. Specifically, the magnetic field that envelops the ring after the passage of the forward shock serves to limit the growth of Kelvin-Helmholtz instabilities at the ring's boundary. Since these instabilities are responsible for the fragmentation and eventual destruction of the ring, the presence of the magnetic field allows the ring to persist for a longer duration. Following \cite{2019A&A...622A..73O}, the ambient magnetic field was modeled as a Parker spiral \citep{1958ApJ...128..664P}, with a strength of $B \approx 1\,\mu$G at the inner edge of the nebula ($r_{\rm HII} = 0.08$~pc), consistent with observations \citep{2018ApJ...861L...9Z}. This field stabilizes CSM inhomogeneities (e.g., the clumpy equatorial ring) by suppressing hydrodynamic instabilities that would develop at their boundaries \citep{2008ApJ...678..274O}, without significantly altering the blast wave evolution.

Following the approach used in the SN model (\citealt{2020ApJ...888..111O}), a re-mapping technique was employed to track the expansion of the SNR. The initial computational domain spanned $6 \times 10^{14}$~cm in all directions, consisting of two regions: a uniform high-resolution grid ($512^3$ zones) focused on the metal-rich ejecta and a non-uniform grid (with an additional 768 zones per direction) covering the remaining domain. This setup provided resolutions ranging from $\approx 2.9 \times 10^{11}$~cm to $\approx 5.8 \times 10^{11}$~cm. As the shock expanded, the domain was iteratively extended, re-mapping physical quantities to the new grid. In Paper I, over 50 years of evolution, approximately 49 re-mappings were performed, resulting in a final domain of $2.6 \times 10^{18}$~cm and a spatial resolution of $\approx 1.3 \times 10^{15}$~cm in the remnant interior. In the present study, we extended simulations B18.3 and B18.3-Dec to 5000 years post-SN, when the remnant reaches a radius of 13 pc, and performed 70 re-mappings. Boundary conditions were maintained at pre-SN CSM values throughout the evolution.

\subsection{Synthesis of X-ray emission}

From the model results, we synthesized the thermal X-ray emission produced by the interaction of the blast wave with the surrounding nebula, following the approach described in previous studies (\citealt{2015ApJ...810..168O, 2024ApJ...977..118O, 2019NatAs...3..236M}). In the original simulation domain, the dense equatorial ring lies in the $[x,y]$ plane. To align the ring’s orientation with that observed in \sna, as inferred from optical observations \citep{2005ApJS..159...60S}, we rotated the domain around the three axes by $i_{\rm x} = 41^\circ$, $i_{\rm y} = -8^\circ$, and $i_{\rm z} = -9^\circ$. After applying these rotations, the symmetry axis of the bipolar explosion is oriented at an angle of $40^{\circ}$ from the line-of-sight (LoS), consistent with the propagation direction of two Ni-rich clumps ejected during the explosion \citep[e.g.,][]{1995A&A...295..129U, 1997ApJ...486.1026N, 2000ApJS..127..141N}. These clumps move in opposite directions at angles ranging from $31^{\circ}$ and $45^{\circ}$ from the LoS \citep[e.g.,][]{1995A&A...295..129U, 2002ApJ...579..671W}. In this configuration, the observer's  LoS corresponds to the negative $y$-axis, ensuring a consistent perspective with the observed morphology and kinematics of the system (see Fig.~9 in \citealt{2020A&A...636A..22O}).

For each cell in the spatial domain, we calculated key physical properties relevant to the X-ray emission. The maximum ionization age is given by $\tau_j = n_{\rm e,j} \Delta t_{\rm j}$, where $n_{\rm e,j}$ is the electron number density and $\Delta t_{\rm j}$ is the time elapsed since the plasma in the cell was shocked. The emission measure is defined as EM$_{\rm j} = n_{\rm e,j} n_{\rm Z,j} V_{\rm j}$, where $n_{\rm Z,j}$ is the ion number density and $V_{\rm j}$ is the cell volume. The electron temperature, $T_{\rm e,j}$, is initially set to $kT = 0.3$ keV at the shock front due to instantaneous heating by lower hybrid waves (\citealt{2007ApJ...654L..69G, 2015ApJ...810..168O}) and is subsequently determined based on the ion temperature, plasma density, and $\Delta t_{\rm j}$, assuming Coulomb collisions.

The X-ray emission in the $[0.1, 10]$ keV band was synthesized based on the values of $\tau_{\rm j}$, EM$_{\rm j}$, and $T_{\rm e,j}$, assuming a source distance of $D = 51.4$ kpc (\citealt{1999IAUS..190..549P}). We employed the non-equilibrium ionization (NEI) emission model VPSHOCK, available in the XSPEC package, using atomic data from ATOMDB (\citealt{Smith2001ApJ}). In this way, the final ion fraction distribution per species comes out as a result from the XSPEC model. Metal abundances for the CSM were taken from deep \textit{Chandra} observations of \sna\ (\citealt{2009ApJ...692.1190Z}), while those for the ejecta were derived from the simulated evolution of the post-core-collapse ejecta. To account for interstellar photoelectric absorption, we applied a column density of $N_{\rm H} = 2.35 \times 10^{21}$ cm$^{-2}$ (\citealt{2006ApJ...646.1001P}) to the X-ray spectrum from each computational cell. The spectra were then convolved with the instrumental response of XRISM Resolve for a comparison with new observations collected with this instrument\footnote{\sna\ was observed on 2024-06-17 for a total of  348866 s (OBS ID 300021010) during the performance verification phase.}.

\section{Results}
\label{sec:result}

A detailed description of the evolution from the SN event to the fully developed remnant at 50 years is provided in two previous studies. \cite{2020ApJ...888..111O} analyzed the SN blast wave’s evolution, from shock revival to breakout, exploring progenitor models (Sect.~\ref{sec:stellar_mod}) and parameterizing an aspherical, bipolar SN explosion. Our study identified a binary merger progenitor (\citealt{2018MNRAS.473L.101U}) with an asymmetric, highly collimated explosion as the best match to observations, particularly in reproducing the [Fe II] line profiles detected 1–2 years after the SN and the $^{56}$Ni mass, providing insights into the explosion mechanism and progenitor’s final stages.

In Paper I, we extended the blast wave evolution to 50 years, tracing its transition to the SNR phase as it interacts with the inhomogeneous CSM. By exploring SNR models and progenitor scenarios, we identified a CSM configuration that best matches key observables. Our preferred model (B18.3; Table~\ref{Tab:model}) reproduces the $^{44}$Ti decay line profile observed by NuSTAR $\sim 25$ years after the SN (\citealt{2015Sci...348..670B}), and the spatial distribution of SiO and CO (ALMA; \citealt{2017ApJ...842L..24A}). Molecular formation calculations \citep{2024ApJS..271...33O} based on the B18.3 model (b18.3-high; \citealt{2020ApJ...888..111O}) further support the consistency with observations. The model also matches the remnant’s X-ray morphology, light curves, and spectra from 1990 to 2016 (\citealt{2006A&A...460..811H, 2012A&A...548L...3M, 2013ApJ...764...11H, 2016ApJ...829...40F}), which helped reconstruct the CSM structure, determine the optimal bipolar explosion orientation with respect to the LoS, and reinforce the binary merger scenario as the most consistent progenitor model.

Although model B18.3 was constrained using observations collected before 2016, it has accurately predicted the subsequent remnant’s X-ray evolution. It reproduced the soft and hard X-ray light curves up to the present day and captured changes in the expansion rate as the blast wave interacts with the inhomogeneous CSM (\citealt{2021ApJ...916...41S, 2021ApJ...922..140R, 2024ApJ...966..147R}). Notably, it also anticipated the emergence of dominant emission from the outermost shocked ejecta around 2021, a prediction recently confirmed by XMM-Newton (\citealt{2025ApJ...981...26S}) observations.

Here, we investigate the structure of both unshocked and shocked ejecta in \sna, evaluating the impact of Ni-bubble effects, by comparing our preferred SNR models (B18.3 and B18.3-Dec) with recent JWST observations to identify potential discrepancies. Identifying mismatches between simulations and observations will guide improvements in the model, shedding light on potentially overlooked physics and enabling a more accurate reconstruction of the remnant's structure, dynamics, and underlying processes. Additionally, we make predictions for XRISM observations, evaluating the detectability of shocked ejecta and outlining key ejecta diagnostics that will soon be testable with XRISM’s high spectral-resolution data. Finally, we provide predictions for the remnant's future evolution, offering insights into its long-term behavior and evolution over the coming millennia. These predictions may help identify features in present-day evolved remnants that reveal early interactions with an inhomogeneous CSM similar to that of \sna.

\subsection{Impact of Ni-bubble on the ejecta properties}
\label{sec:ejecta_prop}

As an initial step, we explored the impact of the Ni-bubble effect on the dynamical evolution of the ejecta (see also \citealt{2021MNRAS.502.3264G}). In model B18.3, the ejecta are characterized by an extended H and He envelope, with the large-scale asymmetry of the bipolar SN explosion reflected in the distribution of innermost metal-rich ejecta. Notably, the Fe-rich ejecta exhibit a distinctive morphology, with two prominent clumps: one propagating northward toward the observer and the other moving southward, away from the observer (see Fig.~4 in Paper I). These structures result from the explosion geometry and the subsequent hydrodynamic interactions, which shape the overall ejecta distribution.

\begin{figure}
   \centering
   \includegraphics[width=0.47\textwidth]{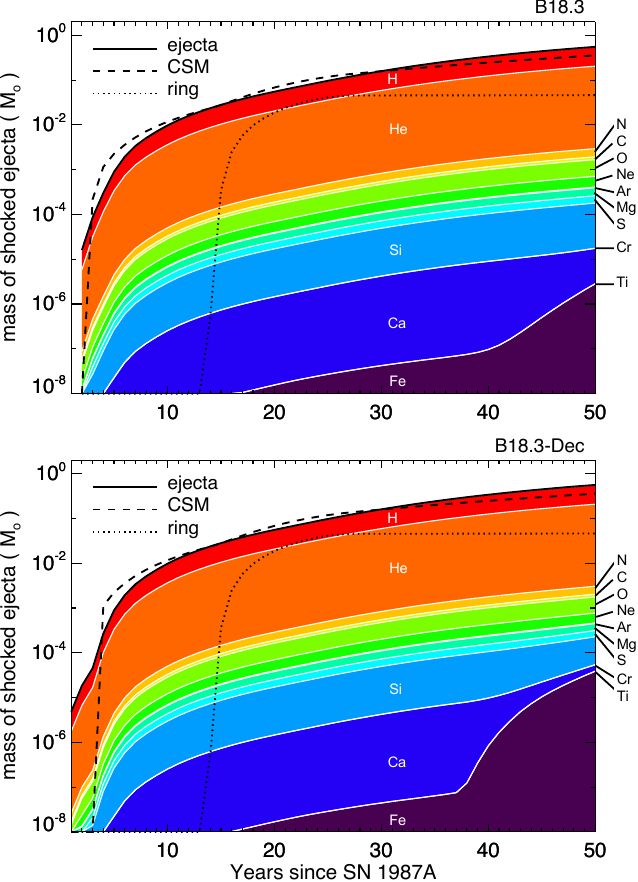}
   \caption{Evolution of the shocked ejecta mass (in units of $M_{\odot}$), in log scale, for various species, as derived from models B18.3 (top panel) and B18.3-Dec (bottom). Different colors represent specific elements, as indicated by the labels. The black solid, dashed, and dotted lines represent the total shocked ejecta mass, the shocked CSM mass (including the H\,II region and the rings), and the mass of shocked material from the equatorial ring, respectively. Due to their low mass content, Ar, Cr, and Ti are barely visible in the figure.}
   \label{ej_content}%
   \end{figure}

To assess potential signatures of the Ni-bubble effect, we analyzed the mass of shocked species composing the ejecta over time relative to the mass of shocked CSM, considering contributions from both the H\,II region and the equatorial ring. Figure~\ref{ej_content} shows the evolution of shocked ejecta mass for various chemical species during the first 50 years post-explosion, as predicted by models B18.3 (upper panel) and B18.3-Dec (lower panel). The black solid, dashed, and dotted lines represent the total shocked ejecta mass, the shocked CSM mass (including the H\,II region and the rings), and the mass of shocked material from the equatorial ring, respectively.

   \begin{figure*}
   \centering
   \includegraphics[width=0.88\textwidth]{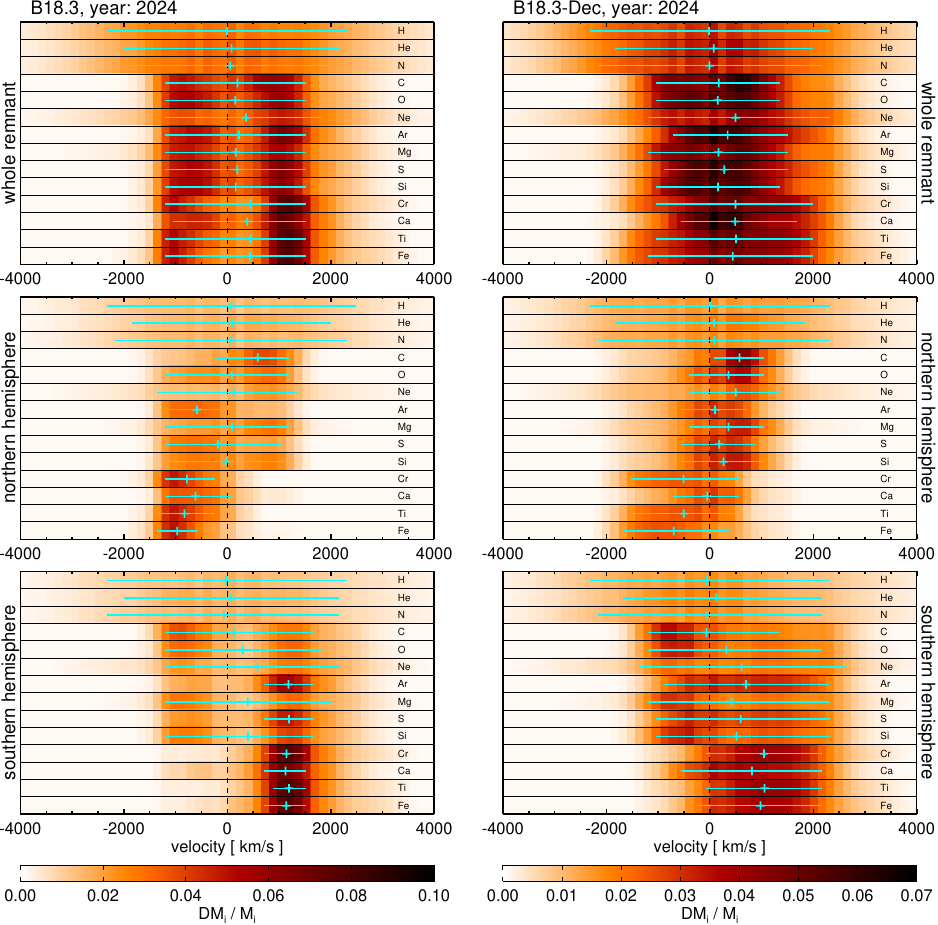}
   \caption{Mass distributions of ejecta species as a function of LoS velocity, derived from models B18.3 (left panels) and B18.3-Dec (right panels) at a remnant age of 37 years (2024). The color bars at the bottom of the figure correspond to their respective columns and are specific to the model represented in each column. The upper panels show the velocity distributions across the entire remnant, while the middle and lower panels correspond to the northern and southern hemispheres, respectively. Crosses mark the mass-weighted average LoS velocity for each species, while the horizontal lines indicate the velocity range where the ejecta mass fraction exceeds 50\% of the peak mass fraction.}
   \label{vel_prof_2024}%
   \end{figure*}

The evolution of shocked material in both models follows three distinct phases (also discussed in Paper I). During the first $\sim 2-3$ years post-explosion, most of the shocked material originates from the outermost ejecta, which expand nearly freely through the low-density wind of the BSG progenitor. In this phase, the effects of the Ni-bubble manifest primarily in the unshocked ejecta, while the outermost ejecta remain largely unaffected. This is driven by the inflation of initially Ni-rich material due to local energy deposition from the radioactive decay of $^{56}$Ni to $^{56}$Co, and subsequently to $^{56}$Fe (see \citealt{2021A&A...645A..66O, 2021MNRAS.502.3264G} for a detailed discussion of the effects of Ni-bubble). As a result, the Fe-rich ejecta in model B18.3-Dec appear more extended than in model B18.3, while the outermost ejecta layers remain essentially unchanged from those in model B18.3. This additional expansion in model B18.3-Dec displaces the surrounding ejecta enriched in heavy and intermediate-mass elements, eventually promoting mixing between chemically distinct layers and leading to a more complex and stratified structure in the innermost regions of the remnant. Approximately one year after shock breakout, the energy input from radioactive decay becomes negligible, and the ejecta in both models continue to expand ballistically.

The second phase begins when the SN blast wave encounters the surrounding H\,II region, leading to a rapid increase in shocked CSM mass, which dominates until $\sim 30$ years (see dashed line in Fig.~\ref{ej_content}). During this period, the shocked equatorial ring plays a key role in maintaining a higher total shocked CSM mass than shocked ejecta (as indicated by the dashed line exceeding the solid line in the figure). The third phase, starting around 2018 according to our models, sees the shocked ejecta mass surpass the shocked CSM, becoming the dominant component. This transition aligns with the findings from the analysis of synthetic X-ray light curves based on model B18.3 in Paper I and is further supported by the comparison between synthetic and observed X-ray light curves \citep{2024ApJ...966..147R} and from the analysis of the distribution of emission measure inferred from XMM-Newton observations (\citealt{2025ApJ...981...26S}).

The colors in Fig.~\ref{ej_content} represent the chemical composition of the shocked ejecta. As expected, H and He dominate during the whole evolution, originating from the outermost progenitor layers interacting with the reverse shock. This is consistent with the presence of a massive H and He envelope in the progenitor of \sna\ (see \citealt{2018MNRAS.473L.101U, 2020ApJ...888..111O}). Over time, the shocked ejecta become increasingly enriched in intermediate-mass and Fe-group elements as deeper layers are processed by the reverse shock.

The impact of the Ni-bubble effect is most evident in the evolution of shocked Fe (see Fig.~\ref{ej_content}). While the evolution of other species remains largely similar between the two models, the interaction of pure Fe ejecta with the reverse shock progresses at different rates in the two models. In model B18.3-Dec, the mass of shocked Fe begins to rise sharply at 38 years (2025), while in model B18.3, this increase is delayed until 41 years (2028) and occurs more gradually. This sharp rise marks the moment when the bulk of the pure Fe-rich ejecta, concentrated in the two clumps propagating in opposite directions, encounters the reverse shock, leading to enhanced mixing with other species.

We further investigated the impact of the Ni-bubble effect on the dynamics of the ejecta by analyzing the asymmetries in the LoS velocity distributions of various species, and computing their mass distributions as a function of LoS velocity. Figure~\ref{vel_prof_2024} shows these distributions for models B18.3 and B18.3-Dec at a remnant age of 37 years (2024), the first epoch observed with XRISM (OBS ID 300021010). However, we found that the profiles remain largely consistent in the near future, indicating that the analysis is applicable to ejecta observed from 2024 through 2037. In the figure, $D M_{\rm i}$ represents the mass of the $i$-th element within the velocity range $[u; u + \Delta u]$, normalized by the total mass $M_{\rm i}$ of that element. The velocity bins are 200 km s$^{-1}$ wide, providing a detailed view of the velocity distribution across the remnant.

The upper panels of Fig.~\ref{vel_prof_2024} show that, in both models, light elements (H, He, and N) maintain mass-weighted average LoS velocities near zero across the entire remnant. Their LoS velocity distributions are broad, with ejecta containing more than 50\% of the peak mass fraction spanning from approximately $-2000$ to $2000$ km s$^{-1}$, and extending to even higher velocities in the wings. This behavior is consistent with the location of these light elements in the outermost ejecta layers, which are relatively unaffected by the asymmetries induced by the explosion.

For intermediate-mass elements (C, O, Ne, Ar, Mg, S, and Si) and heavier elements (Cr, Ca, Ti, and Fe), a slight redshift is seen in the mass-weighted average LoS velocities. These elements exhibit mass-weighted average velocities ranging from $\sim 300$ km s$^{-1}$ for intermediate-mass elements to $\sim 500$ km s$^{-1}$ for heavier elements. Their velocity distributions show a moderate spread, generally between -1000 and 1500 km s$^{-1}$. A key distinction between the models lies in the shape of these velocity distributions. In model B18.3, two distinct peaks (one redshifted and one blueshifted) are present, whereas in model B18.3-Dec, these peaks merge into a single dominant redshifted peak. The peaks in B18.3 arise from the two Fe-rich ejecta clumps formed by the bipolar explosion (see Fig.~4 in Paper I) and influencing the surrounding ejecta layers. In contrast, the Ni-bubble effect in model B18.3-Dec enhances the expansion of these Fe-rich clumps, leading to their partial merging and resulting in a broader redshifted peak in the velocity distribution.

A more detailed analysis of the northern and southern hemispheres (middle and lower panels in Fig.~\ref{vel_prof_2024}) reveals that the light elements (H, He, and N) show no significant hemispheric asymmetry, as their velocity distributions remain broad and centered around zero. In contrast, intermediate-mass elements (C, O, Ne, Ar, Mg, S, and Si) exhibit predominantly redshifted distributions in both hemispheres, except for Ar (and marginally S and Si) in model B18.3, which is blueshifted in the northern hemisphere. The mass-weighted velocities of these elements are approximately 300 km s$^{-1}$ in the northern hemisphere, while in the southern hemisphere, they reach values up to 1000 km s$^{-1}$.

The behavior of heavy elements (Cr, Ca, Ti, and Fe) is more striking, as they exhibit a clear hemispheric asymmetry. In the northern hemisphere, these elements are predominantly blueshifted, with velocities ranging from $-500$ to $-1000$ km s$^{-1}$, while in the southern hemisphere, they are redshifted, with velocities ranging from 1000 to 1500 km s$^{-1}$. This pronounced asymmetry highlights the significant role of explosion-induced asymmetries (likely tied to the explosion mechanism itself) in driving the expansion of Fe-rich ejecta and shaping the surrounding layers. The impact of the Ni-bubble effect is most evident in the broader distribution of the mass fractions of these species across LoS velocities. This results in the merging of the two distinct peaks observed in model B18.3 into a single dominant redshifted peak in model B18.3-Dec, further emphasizing the influence of the Ni-bubble on the remnant’s dynamical evolution.

  \begin{figure*}
   \centering
   \includegraphics[width=0.98\textwidth]{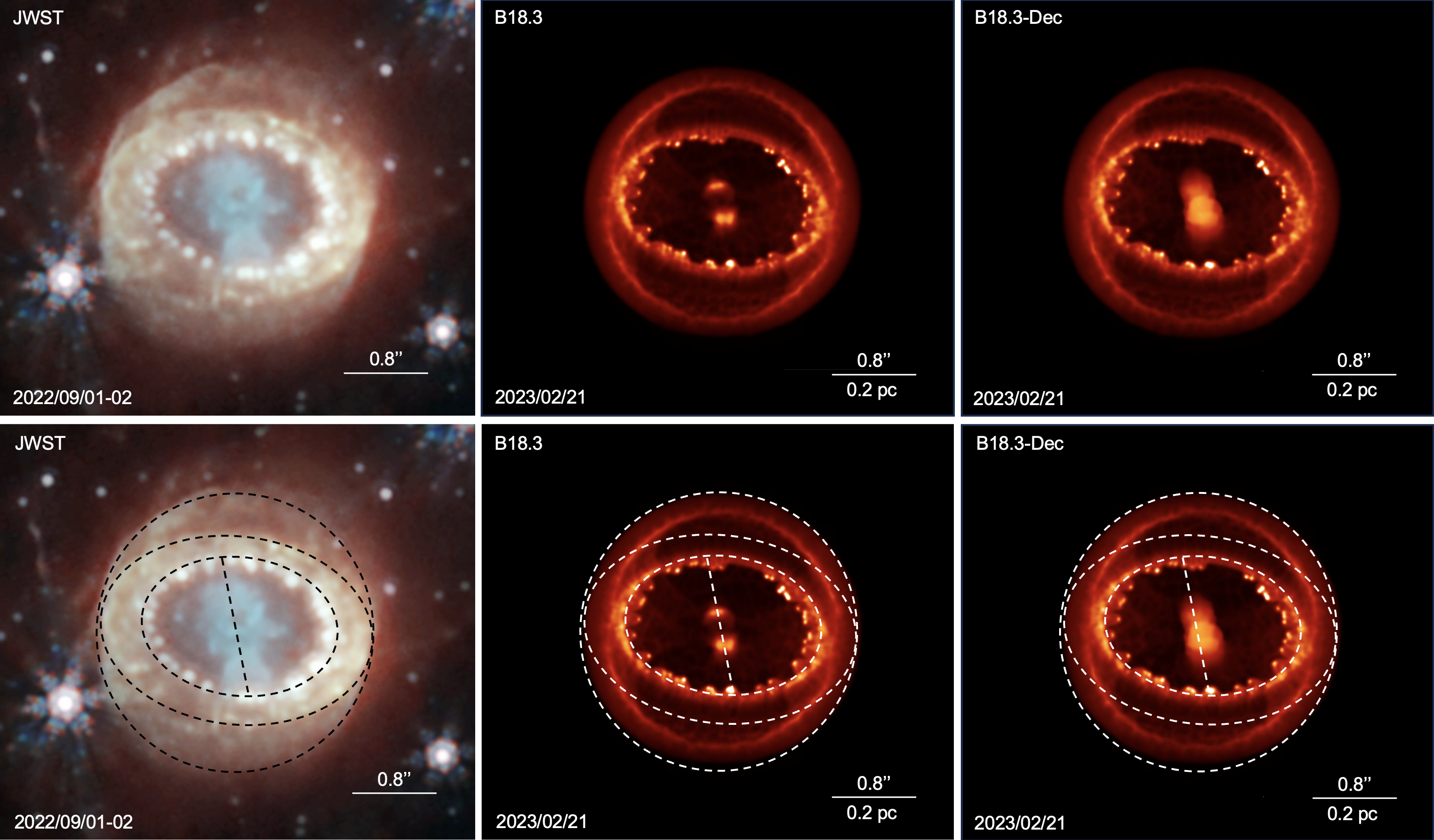}
   \caption{Upper panels: Comparison between JWST observations of \sna\ from September 1–2, 2022 (left column) and the density distribution of the remnant—including both shocked plasma and unshocked Fe-rich ejecta—predicted by models B18.3 (Paper I; center column) and B18.3-Dec (right column) as of February 2023. Lower panels: Same as the upper panels, but with identical ellipsoids overlaid on all images to serve as a reference for comparing the large-scale morphology of the remnant across observations and models. The dashed line at the remnant’s center marks the approximate orientation of the homunculus’ major axis. While the model-generated images reproduce the general structure of the remnant, they are not intended to exactly match the observed appearance. JWST image credits: NASA, ESA, CSA, Mikako Matsuura (Cardiff University), Richard Arendt (NASA-GSFC, UMBC), Claes Fransson (Stockholm University), and Josefin Larsson (KTH).}
   \label{comparison}%
   \end{figure*}

\subsection{Unshocked ejecta: comparison with JWST data}
\label{sec:jwst}

The primary interest in the unshocked ejecta lies in their potential to preserve signatures of the physical processes that governed the SN explosion. Identifying these signatures is essential for constraining the explosion mechanism (e.g., \citealt{2025A&A...696A.108O}). Recent JWST observations of \sna\ provide valuable insights into the properties and spatial distribution of the ejecta. A comparison with our models can further clarify the asymmetries present at the onset of the explosion, offering a deeper understanding of the underlying dynamics.

Figure~\ref{comparison} compares deep near-infrared imaging of \sna\ obtained with JWST/NIRCam on September 1–2, 2022 (\citealt{2024MNRAS.532.3625M}) with 3D volumetric renderings of the density distribution of shocked plasma, as well as the unshocked Fe-rich ejecta, predicted by the two models in Table~\ref{Tab:model}, as of February 2023\footnote{The simulations produced one snapshot per year; the selected snapshot is the one nearest in time to the observational data.}. The upper panels provide an unobstructed comparison between models and observations. The lower panels show the same images with identical ellipsoids overlaid, providing a common reference frame to facilitate the comparison of the remnant’s large-scale morphology between the JWST observations and the simulations. The dashed line at the center of the remnant indicates the approximate projected inclination of the elongated distribution of the inner ejecta, which is visible in cyan in the JWST image.

It is important to note that while the model-derived images combine the distributions of shocked plasma and unshocked Fe-rich ejecta to create an analog of the JWST observations (left panels of Fig.~\ref{comparison}), they are not intended to precisely replicate the observational data. Instead, these images illustrate the remnant’s structure and emphasize key morphological features. A direct match between model-derived images and observations would require detailed emission modeling in the specific bands observed by JWST. For example, the infrared image exhibits a "halo" around the brightest structures due to radiative ionization from photons, an effect not accounted for in the sharp density maps. Incorporating such a halo around density structures could make the model images appear more extended and better resemble the observations. Despite this limitation, the strong similarity between the model-derived images and JWST observations underscores the model’s ability to reproduce the global structure and dimensions of the remnant, capturing many of the key features observed in \sna.

However, as discussed below, some discrepancies remain between the models and observations, highlighting the need for further refinements to achieve a better match. The remnant's morphology is dominated by the bright, shocked equatorial ring, which appears in projection as a slightly inclined, clumpy ellipsoid. Observations indicate that the ring is more radially extended than predicted by the models, likely due to the sharper appearance of the density maps, as discussed above. Additionally, JWST observations have revealed the emergence of new bright knots (e.g., \citealt{2024MNRAS.532.3625M}) that are not present in the models. These structures likely form as the blast wave propagates through the CSM, unveiling previously undetectable inhomogeneities (and, therefore, not included in the model developed before these observations). Their appearance suggests that the CSM is more complex than currently modeled, possibly containing finer-scale density variations or unresolved clumps that can influence the remnant's evolution and emission properties.

The models successfully reproduce the diffuse emission observed beyond the equatorial ring, as indicated by the largest ellipsoid in Fig.~\ref{comparison}. This emission appears more elongated in the north-south direction, consistent with the blast wave propagating through the hourglass-shaped cavity carved by the fast wind from the BSG progenitor. However, differences in the fine structure and distribution of the shocked material suggest that additional physical processes, such as small-scale instabilities or variations in the CSM, may play a role in shaping the observed remnant.

The most striking discrepancies between the models and observations arise in the distribution of the innermost ejecta, which are rich in heavy elements such as Si, S, Ti, and Fe. In the models, these ejecta exhibit an elongated morphology, consistent with the observed structure in \sna, as a consequence of the bipolar explosion implemented in the SN model. As expected, the structure of the innermost Fe-rich ejecta differs depending on whether Ni-bubble effects are considered during the first year of evolution. In model B18.3-Dec (right column in Fig.~\ref{comparison}), the Fe-rich ejecta appear more expanded than in model B18.3 (center column), reflecting the additional pressure-driven expansion induced by the decay of radioactive $^{56}$Ni (see also \citealt{2021MNRAS.502.3264G}).

The orientation of the modeled elongated ejecta also aligns well with observations (see the reference dashed line at the center of the lower panels in Fig.~\ref{comparison}). However, the modeled metal-rich ejecta appear significantly less extended than those observed (compare the modeled distributions with the observed one in the left column of the figure). This discrepancy is evident even in model B18.3-Dec, suggesting that Ni-bubble effects alone cannot account for the observed degree of expansion. Additional factors, such as a more pronounced asymmetry in the SN explosion, may help produce the observed ejecta distribution. 

\subsubsection{3D structure of reverse shock and Fe-rich ejecta}

The structure of the innermost ejecta as observed with JWST was extensively analyzed by \cite{2023ApJ...949L..27L}, who reconstructed 3D emissivity maps of the [Fe I] 1.443 $\mu$m and He I 1.083 $\mu$m line. Their study revealed a highly asymmetric distribution of Fe-rich material, characterized by a distinct broken-dipole morphology. More specifically, the ejecta structure is dominated by two large, spatially separated clumps of Fe-rich ejecta expanding at velocities of approximately 2300~km~s$^{-1}$, highlighting significant deviations from spherical symmetry (see Fig.~3 in \citealt{2023ApJ...949L..27L}). These findings provide crucial constraints on the explosion geometry and suggest that the metal-rich ejecta experienced strong directional asymmetries during the SN event.

   \begin{figure*}
   \centering
   \includegraphics[width=0.9\textwidth]{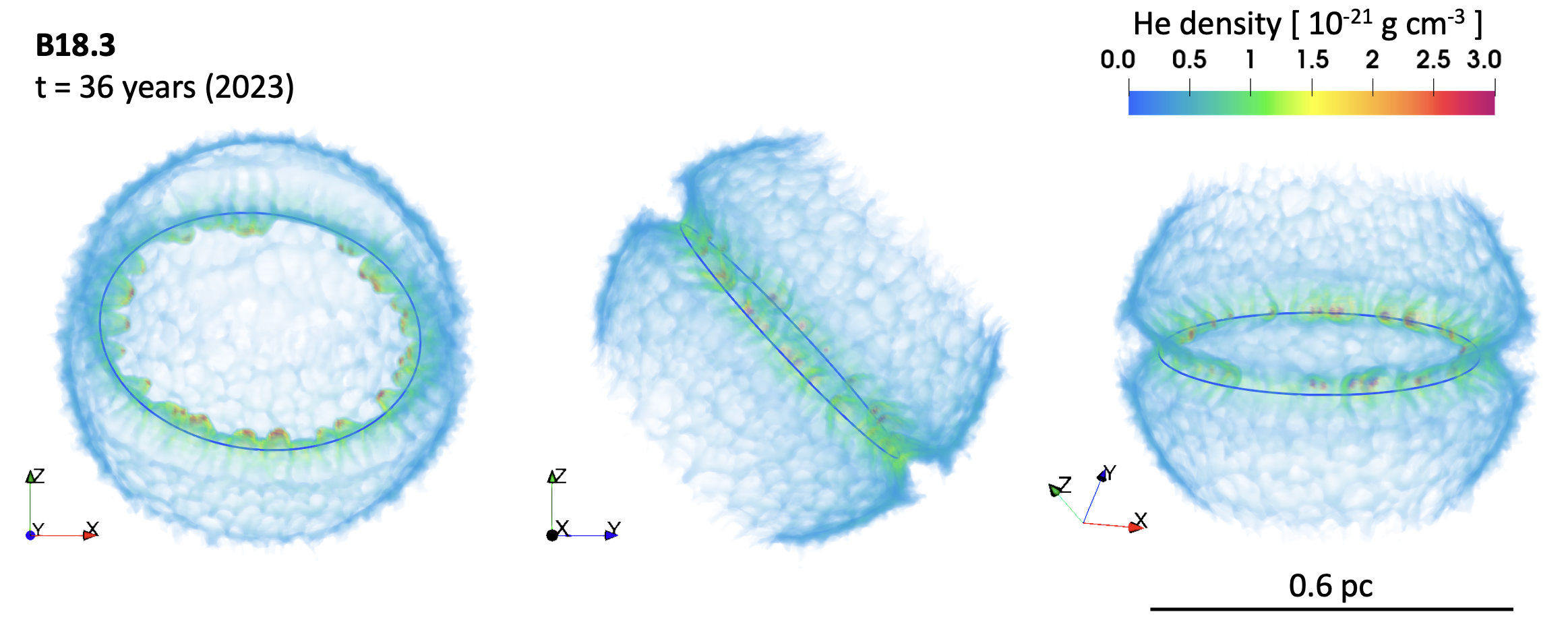}
   \caption{3D volume rendering of He-rich ejecta near the reverse shock, with opacity scaled to reflect density variations, shown from different viewing angles. The visualization is based on model B18.3 at $t=36$ years after the SN (corresponding to February 2023). The color bar in the upper right indicates the He density. The blue circle indicates the nominal position of the equatorial ring. Earth’s vantage point corresponds to the negative $y$-axis.}
   \label{He_distrib}%
   \end{figure*}
   \begin{figure*}
   \centering
   \includegraphics[width=\textwidth]{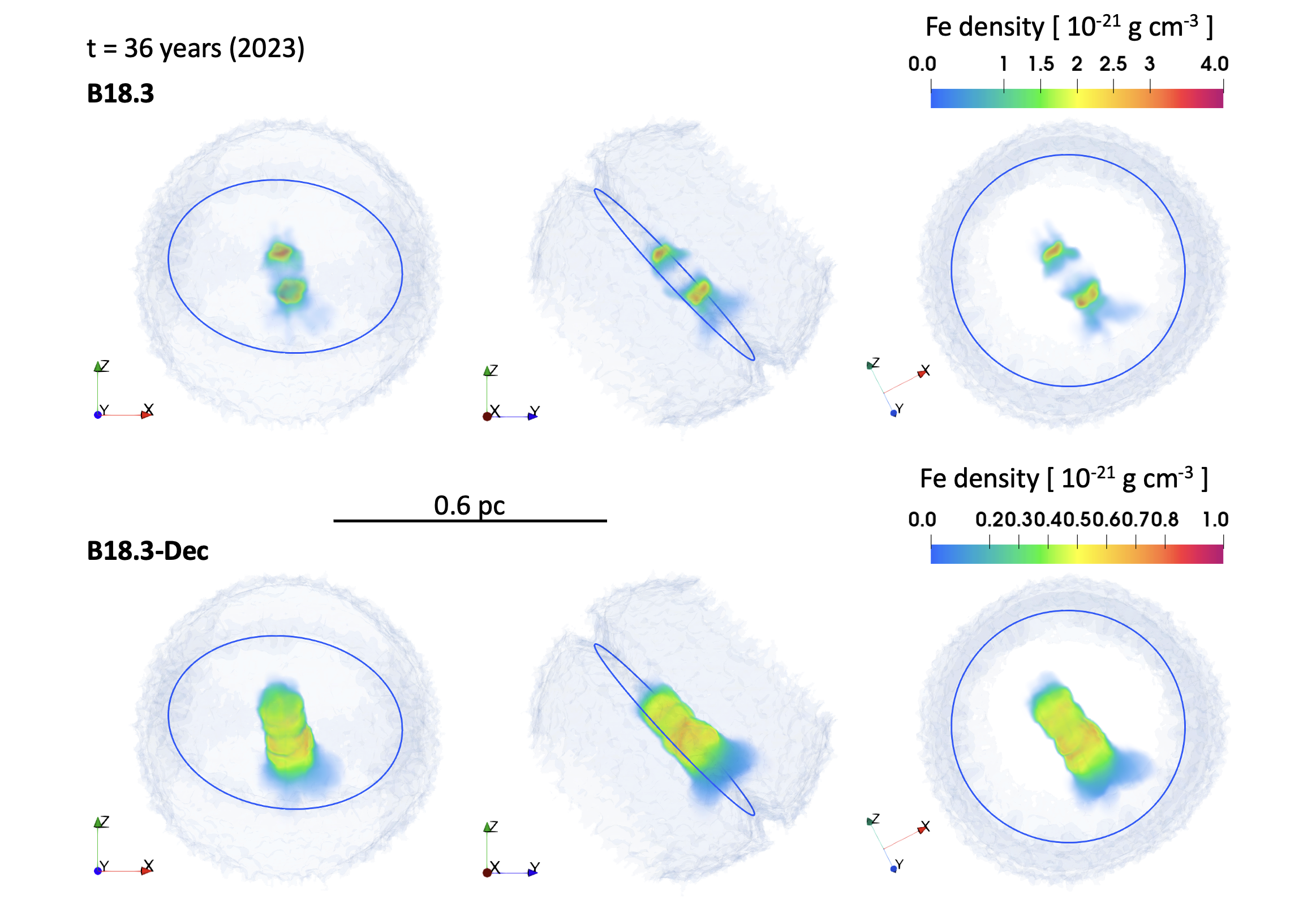}
   \caption{3D volume rendering of Fe-rich ejecta, with opacity scaled to reflect density variations, shown from different viewing angles. The visualization is based on model B18.3 (upper panels) and B18.3-Dec (lower panels) at $t=36$ years after the SN (corresponding to February 2023). The color bars on the right indicate the Fe density for each case. The blue circle marks the nominal position of the equatorial ring, while the semi-transparent diffuse structure represents the distribution of He-rich ejecta near the reverse shock, as shown in Fig.~\ref{He_distrib}. Earth’s vantage point corresponds to the negative $y$-axis.}
   \label{Fe_distrib}%
   \end{figure*}

The distribution of Fe-rich ejecta in relation to the reverse shock in our models is shown in Figs.~\ref{He_distrib} and \ref{Fe_distrib}. The reverse shock's morphology is inferred from the distribution of freshly shocked He-rich ejecta (see Fig.~\ref{He_distrib}). We found that both models analyzed exhibit the same overall structure of the reverse shock, as shown in the figure for model B18.3. This similarity arises because the Ni-bubble effects do not significantly impact the dynamics of the outermost ejecta, particularly the H and He envelope interacting with the reverse shock at the time of the JWST observations. Interestingly, the modeled structure of the reverse shock closely resembles that reconstructed from the analysis of He I 1.083 $\mu$m emission in JWST data (compare Fig.~\ref{He_distrib} with Fig.~6 in \citealt{2023ApJ...949L..27L}).

The reverse shock is only partially visible in Fig.~\ref{He_distrib}, extending up to latitudes of approximately $45^{\circ}$ above and below the equatorial ring, with the He density gradually decreasing toward higher latitudes. According to our models, the visibility of the reverse shock depends on the density of the shocked ejecta. In polar directions, where the blast wave expands almost freely through the tenuous wind of the BSG progenitor, the reverse shock penetrates less into the ejecta, affecting lower-density regions. As a result, it is less prominent in these directions. In contrast, the reverse shock is most evident in regions where the forward shock propagates through the H\,II region. Consequently, the overall geometry of the reverse shock is nearly spherical, except in regions where the forward shock expands rapidly at the poles, resulting in a lack of visible structure, and in a bottleneck feature at the interaction site with the dense equatorial ring. In this region, the reverse shock appears more recessed due to the remnant’s interaction with the ring, while the He density is enhanced by the strong reflected shock (see Fig.~\ref{He_distrib}). We conclude that the similar morphology of the reverse shock observed in JWST data provides direct evidence of interaction with a CSM whose geometry and density distribution closely match the predictions of our models.

Figure~\ref{Fe_distrib} illustrates the distribution of Fe-rich ejecta in relation to the reverse shock and the equatorial ring for the two models analyzed. As discussed above, model B18.3 predicts the presence of two distinct Fe-rich clumps resulting from the bipolar explosion (see upper panels in Fig.~\ref{Fe_distrib}), while model B18.3-Dec predicts a continuous, elongated structure formed by the merging of the original Fe-rich clumps (lower panels in the figure). The position and propagation direction of these structures show a rough agreement with observations. However, the modeled Fe-rich ejecta appear more interior to the remnant than observed, suggesting that additional factors should be considered. For instance, the explosion asymmetry in the model may not be sufficient to fully reproduce the large-scale asymmetry observed in \sna. Other potential contributors to this discrepancy include the structure of the progenitor star at the time of collapse or the degree of mixing and clumping in the ejecta, which affects their expansion and deceleration. 

   \begin{figure*}
   \centering
   \includegraphics[width=0.95\textwidth]{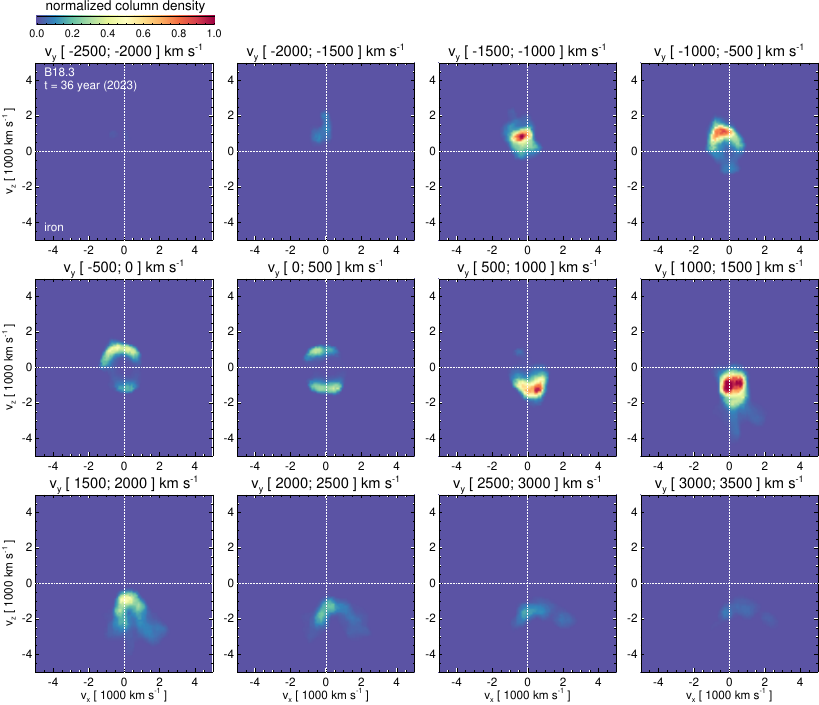}
   \caption{Distribution of Fe-rich ejecta from model B18.3 at $t \approx 36$~years after core collapse (corresponding to February 2023) as a function of velocity along the $y$-axis (LoS), ranging from $v_{\rm y} = -2500$~km~s$^{-1}$ (top left panel, near side of the remnant) to $v_{\rm y} = 3500$~km~s$^{-1}$ (bottom right panel, far side of the remnant). Each panel shows the Fe-mass density integrated over a 500~km~s$^{-1}$ velocity interval along $v_{\rm y}$, with the corresponding range labeled above each frame. The color scale (top left of the figure) indicates the normalized column density of Fe. The axes represent velocity components in the plane of the sky: $v_{\rm x}$ (east-west direction) and $v_{\rm z}$ (south-north direction).}
   \label{vel_map_nodec}%
   \end{figure*}

   \begin{figure*}
   \centering
   \includegraphics[width=0.95\textwidth]{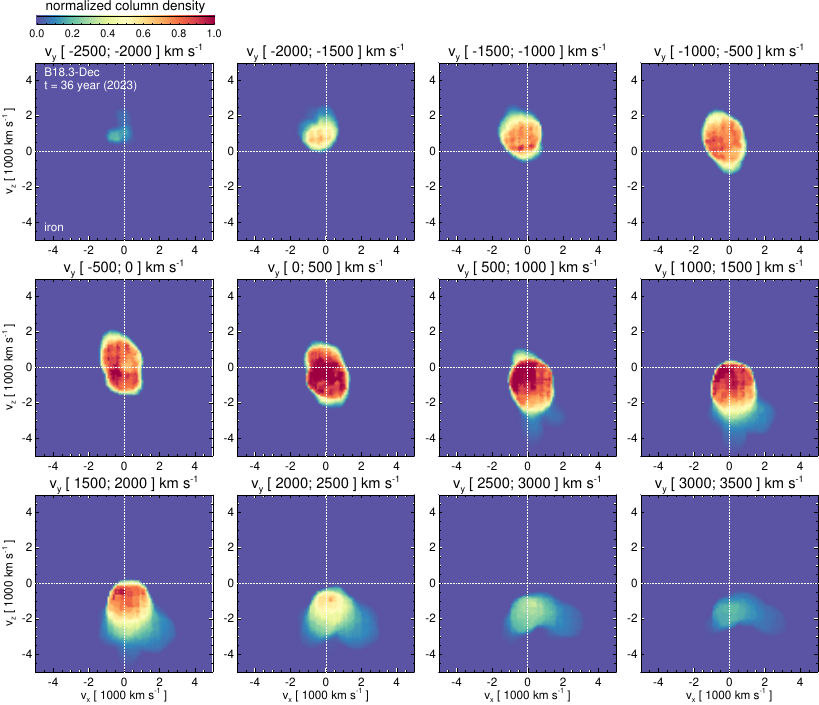}
   \caption{Same as Fig.~\ref{vel_map_nodec}, but for model B18.3-Dec, illustrating the impact of Ni-bubble effects on the distribution and expansion of Fe-rich ejecta. The normalization of the column density differs from that in Fig.~\ref{vel_map_nodec}.}
   \label{vel_map_dec}%
   \end{figure*}

\subsubsection{Dynamical properties of Fe-rich ejecta}

The analysis of the dynamical properties of shocked ejecta presented in Sect.~\ref{sec:ejecta_prop} has revealed significant asymmetries in the velocity distributions of metal-rich ejecta. Similar asymmetries are expected in the unshocked ejecta. To investigate these and enable a more quantitative comparison with JWST observations, we adopted a method that allows for a direct comparison between the model predictions and the observational data. Specifically, we generated column density maps of the modeled Fe-rich ejecta as a function of Doppler shift, providing a detailed visualization of the ejecta's distribution in velocity space. This allows for a direct comparison with the observational data, presented in an analogous format (e.g., Fig.~3 in \citealt{2023ApJ...949L..27L}).

Figures~\ref{vel_map_nodec} and \ref{vel_map_dec} present the distributions of Fe-rich ejecta from models B18.3 and B18.3-Dec, respectively, as a function of velocity along the $y$-axis, which corresponds to the LoS velocity. Negative values indicate blueshifted material (approaching the observer), while positive values correspond to redshifted material (receding from the observer). The spatial scale is mapped onto a velocity scale, where $v_{\rm x}$ represents east-to-west motion, and $v_{\rm z}$ represents south-to-north motion.

Each panel in Figs.~\ref{vel_map_nodec} and \ref{vel_map_dec} displays the integrated ejecta density over 500~km~s$^{-1}$ intervals in $v_{\rm y}$, covering the velocity range $v_{\rm y} = [-2500, 3500]$~km~s$^{-1}$. This representation allows for a detailed examination of how Fe-rich ejecta are distributed along the LoS, providing insights into the 3D structure of the innermost ejecta. Since Earth is positioned along the negative $y$-axis, this approach effectively distinguishes ejecta located on the near side (blueshifted) from those on the far side (redshifted), offering a direct comparison with observational velocity maps (compare with Fig.~3 in \citealt{2023ApJ...949L..27L}).

The maps are qualitatively in agreement with the results from JWST observations, showing two distinct clumps of Fe-rich ejecta. One clump is traveling toward the north, approaching the observer (blueshifted) with a small velocity component eastward (negative $v_{\rm x}$), while the other, more massive clump is traveling south, receding from the observer (redshifted) with a small westward component (positive $v_{\rm x}$). The mass asymmetry between the two clumps originates from the explosion’s asymmetry across the equatorial plane, as established in \citealt{2020ApJ...888..111O} (see also Fig.~2 of Paper I). However, both models predict significantly lower velocities for the two clumps compared to the observed values. 

In model B18.3, the bulk of Fe-rich ejecta exhibit LoS velocities, $v_{\rm y}$, ranging from $-1500$ to $+2500$~km~s$^{-1}$, with a faint tail extending up to $+3500$~km~s$^{-1}$. The velocity components $v_{\rm z}$ range between $-1000$ and $-1500$~km~s$^{-1}$ for the clump propagating to the north, and between $+1000$ and $+1500$~km~s$^{-1}$ for the clump propagating to the south. In total, the two modeled Fe-rich clumps travel at velocities of approximately 1500~km~s$^{-1}$, whereas JWST observations report values of approximately 2300~km~s$^{-1}$ (\citealt{2023ApJ...949L..27L}).

\begin{figure}
   \centering
   \includegraphics[width=0.45\textwidth]{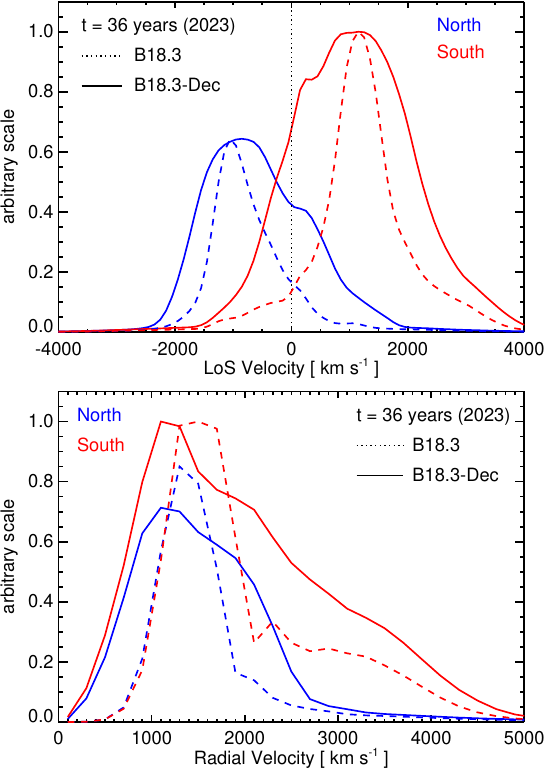}
   \caption{Mass distributions of Fe-rich ejecta as a function of the velocity component along the LoS (upper panel) and the radial velocity (lower panel) at $t \approx 36$~years after core collapse (corresponding to February 2023) for models B18.3 (dashed lines) and B18.3-Dec (solid lines). Red and blue curves represent the results for the southern and northern hemispheres, respectively.}
   \label{Fe_vel_prof}%
   \end{figure}

The inclusion of the Ni-bubble in model B18.3-Dec does not significantly alter the overall results. Its primary effect is the inflation of ejecta initially rich in Ni, leading to more extended Fe-rich clumps (see lower panels in Fig.~\ref{Fe_distrib}). Consequently, while the tails of Fe-rich ejecta can reach velocities up to $\approx 3500$~km~s$^{-1}$, the bulk of the ejecta exhibit velocities similar to those found in model B18.3 (compare Figs.~\ref{vel_map_nodec} and \ref{vel_map_dec}). Another key difference from model B18.3 is that, while the latter shows two clumps connected by a tenuous distribution of Fe-rich material, the inclusion of Ni-bubble effects leads to a significant amount of Fe-rich material between the two clumps, within the velocity range of -500 to 500~km~s$^{-1}$.

The above effect is evident in Fig.~\ref{Fe_vel_prof}, which presents the normalized mass distributions of Fe-rich ejecta as a function of the LoS velocity (upper panel) and radial velocity (lower panel) for both models. The distributions in the two models peak at nearly the same LoS velocities, though the inclusion of the Ni-bubble results in broader distributions in model B18.3-Dec (upper panel), leading to a partial blending of the two peaks (as also noted in Fig.~\ref{vel_prof_2024}). The effects of the Ni-bubble are also evident in the distributions of the radial velocities (lower panel). The broader radial velocity distributions in model B18.3-Dec result in a non-negligible fraction of Fe-rich ejecta reaching velocities above $2000$~km~s$^{-1}$. Additionally, the figure shows that, in both models, the Fe-rich clump in the southern hemisphere is more massive than the one propagating in the northern hemisphere, as indicated by the more prominent red peaks compared to the blue ones. This is in agreement with observations.

It is worth noting that model B18.3-Dec incorporates Ni-bubble effects under the assumption of no $\gamma$-ray leakage from the inner regions of the remnant (see Sect.~\ref{sec:model}). Consequently, the Ni-bubble influence is likely overestimated in this model. A more realistic treatment, accounting for partial $\gamma$-ray escape, would likely result in a Fe-rich ejecta distribution intermediate between those predicted by models B18.3 (no Ni-bubble) and B18.3-Dec (maximum Ni-bubble effect). Furthermore, escaping $\gamma$-rays could contribute to photoionization in the surrounding material, potentially leading to the formation of a halo around emitting structures. Future investigations incorporating improved radiative transport and mixing processes will be crucial for refining these predictions.

\begin{figure}
   \centering
   \includegraphics[width=0.47\textwidth]{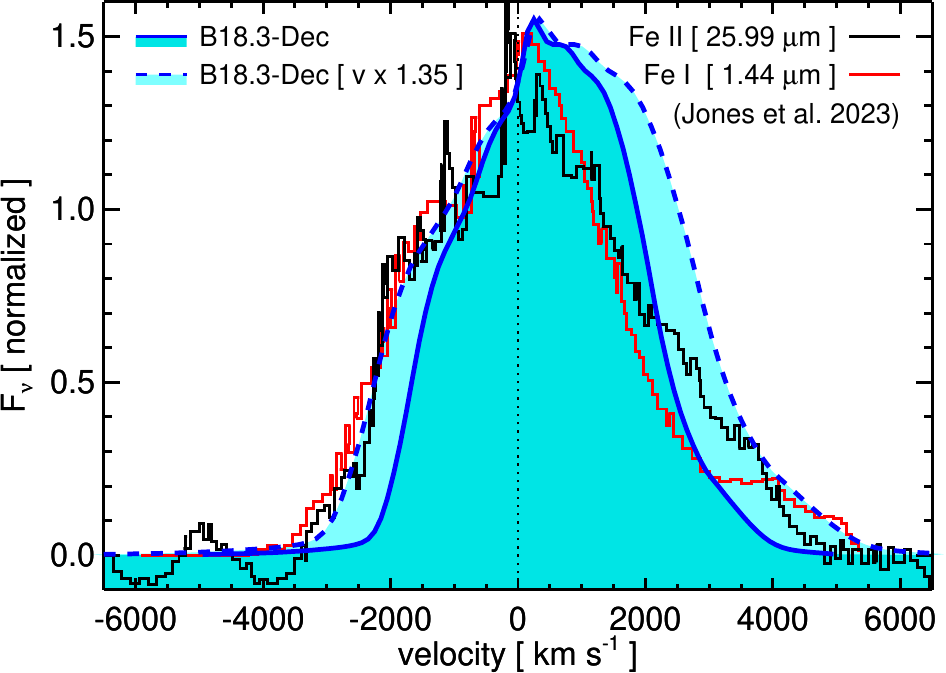}
   \caption{Profiles of continuum-subtracted [Fe II] (25.99 $\mu$m; black line) and [Fe I] (1.44 $\mu$m; red line) emission lines from the inner ejecta of \sna, as observed on 2022 July 16 with JWST (\citealt{2023ApJ...958...95J}). These are compared with the total mass distribution of unshocked Fe-rich ejecta as a function of the LoS velocity, predicted by model B18.3-Dec for February 2023 (solid blue line). The dashed blue line represents the Fe mass distribution with an artificially increased LoS velocity (by 35\%) to better match the JWST observations (\citealt{2023ApJ...949L..27L}). The Fe mass distributions have been normalized to roughly match the peak of [Fe I] line.}
   \label{Fe_lines}%
   \end{figure}

The emission line profiles from the inner Fe-rich ejecta are expected to partially reflect the total mass distribution of unshocked Fe-rich ejecta as a function of the LoS velocity. Figure~\ref{Fe_lines} presents the continuum-subtracted profiles of [Fe II] (25.99 $\mu$m; black line) and [Fe I] (1.44 $\mu$m; red line) emission lines from the inner ejecta of \sna, as observed on 2022 July 16 with JWST (\citealt{2023ApJ...958...95J}). For comparison, the figure also includes the modeled Fe mass distribution profile (solid blue line), derived from model B18.3-Dec by summing the contributions from both the southern and northern hemispheres, as shown in Fig.~\ref{Fe_vel_prof}. As expected, the modeled profile is narrower than the observed emission lines, indicating that the simulated Fe-rich ejecta expand more slowly than observed. Additionally, there is an excess of Fe mass at velocities between 0 and 2000 km~s$^{-1}$, suggesting that the dynamics of the modeled clump receding from the observer differ from those inferred from observations. To further explore this velocity discrepancy, we artificially increased the expansion velocity of the Fe-rich ejecta by 35\% to match the observed values (dashed blue line). This adjustment brings the blueshifted modeled profile into remarkable agreement with observations, but significant discrepancies persist on the redshifted side. This suggests that while the modeled blueshifted clump (moving toward the observer in the northern hemisphere) primarily suffers from an underestimated velocity, the modeled redshifted clump has more fundamental differences in its dynamical properties compared to observations.

The discrepancy in the redshifted clump may be related to another significant mismatch between the models and observations, specifically the propagation direction of the two Fe-rich clumps. JWST observations indicate that the Doppler shifts of the clumps suggest they are not aligned along the same axis passing through the explosion center (\citealt{2023ApJ...949L..27L}). The blueshifted velocity of the northern clump suggests that it lies close to the plane of the equatorial ring, while the smaller redshifted velocity of the southern clump implies that its trajectory is inclined at a larger angle relative to this plane (see also \citealt{2010A&A...517A..51K, 2016ApJ...833..147L}). In contrast, our models predict that both Fe-rich clumps are aligned along the same axis and positioned near the plane of the equatorial ring (see Fig.~\ref{Fe_distrib}). This alignment arises from the idealized bipolar explosion geometry assumed in the initial conditions of the SN model (\citealt{2020ApJ...888..111O}). As a result, while the trajectory of the northern clump is consistent with observations, that of the southern clump deviates from the observed structure.

Interestingly, \cite{2023ApJ...949L..27L} also identified an isolated Fe-rich clump interacting with the reverse shock in the southern hemisphere. This clump is located nearly in the plane of the equatorial ring and closely aligned with the axis passing through the explosion center and the blueshifted Fe-rich clump. This evidence fits well in the general scenario described in our models but suggests a more asymmetric explosion. The misalignment of Fe clumps along a single symmetry axis, combined with discrepancies in the expansion velocity of Fe-rich ejecta, indicates that further refinements in modeling early-time explosion asymmetries and ejecta evolution are necessary to improve agreement between simulations and observations.

   \begin{figure*}
   \centering
   \includegraphics[width=0.95\textwidth]{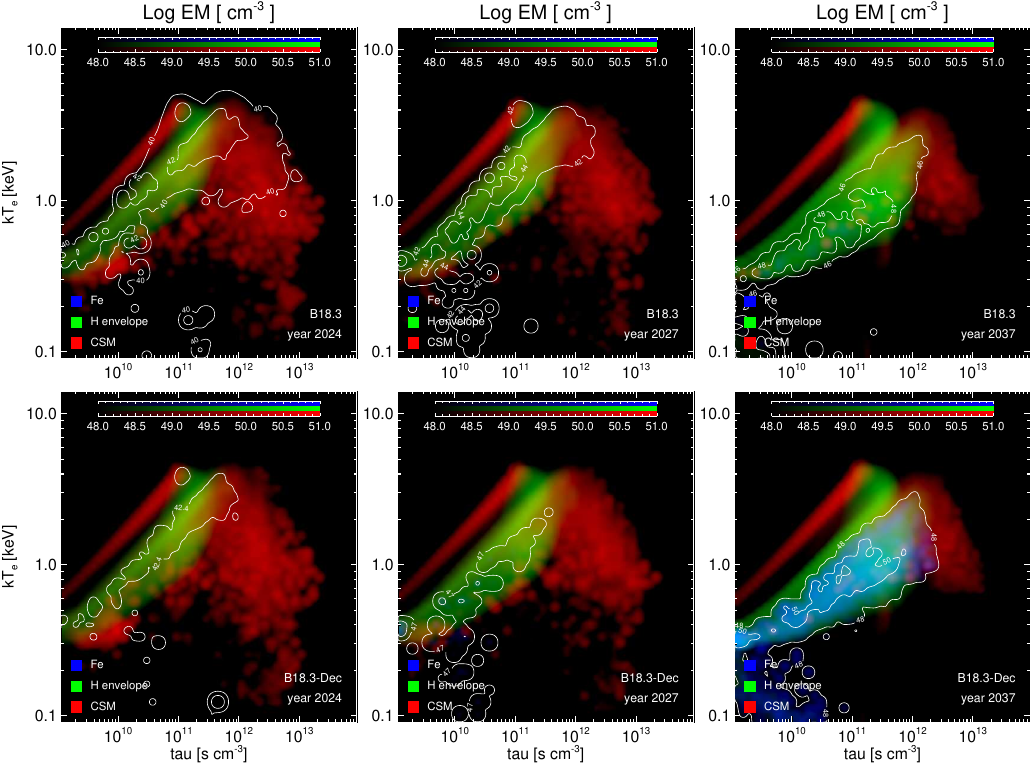}
   \caption{Distributions of EM as a function of electron temperature ($kT_{\rm e}$) and ionization timescale ($\tau = n_{\rm e}t$) at the labeled times for models B18.3 (upper panels) and B18.3-Dec (lower panels). The panels present three-color composite images of the EM distributions, where different colors represent contributions from distinct shocked plasma components: shocked CSM (red), shocked material from the H envelope (green), and shocked Fe-rich ejecta (blue). White contours highlight the regions dominated by Fe.}
   \label{maps_EM}%
   \end{figure*}

\subsection{Shocked ejecta: potential insights from XRISM spectra}

While JWST enables the investigation of the structure of the unshocked innermost ejecta, XRISM spectra allow us to probe the regions where the outermost ejecta interact with the inhomogeneous CSM. These observations thus provide valuable insights into the density distribution and geometry of the CSM, as well as the structure of the outermost ejecta. In particular, constraining the properties of freshly shocked ejecta can offer crucial information about the progenitor star by revealing characteristics of its outer envelope. Additionally, the detection of emerging metal-rich ejecta can place important constraints on explosion asymmetries, as demonstrated by the Fe-rich regions observed in Cas~A \citep{2021A&A...645A..66O}.

According to our models, the contribution of shocked ejecta to the X-ray emission in \sna\ became comparable to that of shocked CSM around 2018 (Paper I; see also Sect.~\ref{sec:ejecta_prop}). In the coming years, the models predict that this contribution will continue to increase, while the emission from the shocked CSM will gradually decline (see Fig.~6 in Paper I). We evaluated, therefore, the observability of shocked ejecta in the X-ray band, providing insights for interpreting current and future observations. In particular, we explored the diagnostic potential of high-spectral-resolution data from the XRISM satellite, identifying key spectral features that trace ejecta properties, shock dynamics, and mixing processes within the remnant.

\subsubsection{Emission measure distribution versus ionization age and electron temperature}

As discussed in Paper I, the X-ray-emitting plasma can be characterized by deriving its EM distribution as a function of electron temperature ($kT_{\rm e}$) and ionization timescale ($\tau$). Figure~\ref{maps_EM} presents the evolution of the EM distribution for models B18.3 and B18.3-Dec from 2024 (the year of the first XRISM observations of \sna) to 2037, covering the next 13 years of evolution. The figure reveals a complex structure of the EM distribution spanning temperatures from 0.1 to $\approx 5$~keV. This distribution suggests that a significant fraction of the emitting plasma deviates from ionization equilibrium, with ionization timescales ranging from $10^9$ to $10^{13}$~s~cm$^{-3}$. It also highlights contributions from both shocked CSM (red) and shocked ejecta (green), along with the specific contribution of shocked Fe (blue). 

The distributions in both models are similar, with the ejecta and CSM components generally remaining spatially distinct, reflecting their differing thermal and ionization histories (see Fig.~\ref{maps_EM}). The shocked CSM, which surrounds the reverse-shocked ejecta (green/blue region in the figure), exhibits different characteristics depending on its environment: in the dense equatorial ring, it typically has longer ionization timescales (clumpy red structure at $\tau > 10^{12}$~s~cm$^{-3}$ in Fig.~\ref{maps_EM}), whereas in the H\,II region, it shows shorter ionization timescales (red region with $\tau < 10^{11}$~s~cm$^{-3}$). This difference primarily arises from the significantly higher densities of the equatorial ring compared to the surrounding H\,II region. The temperature of the shocked CSM, initially distributed over a broad range from 0.1 to 5 keV in 2024 (left panels in Fig.~\ref{maps_EM}), gradually increases over time. By 2037, it exceeds 0.5 keV, driven by the progressive thermalization of electrons with ions (right panels). Notably, the electron temperature exhibits an inverse relationship with the ionization timescale in the shocked ring. This correlation reflects the presence of multiple CSM clumps with varying densities when the shocked plasma approaches thermal equilibrium: denser clumps have a higher ionization timescale and a lower electron temperature, while less dense clumps exhibit a lower ionization timescale and a higher electron temperature.

The shocked ejecta (green/blue region in Fig.~\ref{maps_EM}) exhibit a broad distribution of temperatures and ionization ages, positioned between the two CSM components. In general, the EM distribution of the ejecta shows a gradual expansion toward higher ionization timescales ($\tau$), while the range of temperatures remains nearly unchanged. A more significant evolution occurs in the chemical composition of the shocked ejecta, which become increasingly enriched in metals as the innermost layers begin to interact with the reverse shock. A notable example is Fe (highlighted by the contours and the blue region in the figure), which progressively contributes more to the EM distribution over time. This effect is particularly pronounced in model B18.3-Dec, where the expansion of Fe-rich ejecta is enhanced by the action of the Ni-bubble. As a result, the contribution of shocked Fe becomes distinctly visible in 2037, as evidenced by the prominent blue region in the lower right panel of Fig.~\ref{maps_EM}.

\subsubsection{Dynamical properties of the shocked ejecta}
\label{sec:dyn_sh_ejecta}

In Sect.~\ref{sec:jwst}, the comparison between our models and recent JWST observations revealed that the modeled Fe-rich clumps are slower and located deeper within the ejecta than observed. This suggests that, in our simulations, the interaction between these clumps and the reverse shock occurs later than in \sna. In fact, \cite{2023ApJ...949L..27L} have provided evidence of the first interaction between Fe-rich ejecta and the reverse shock in the southern hemisphere, indicating that this process was already underway in 2022. In contrast, model B18.3-Dec predicts this interaction to begin around 2027, implying a delay of at least five years. As discussed in  Sect.~\ref{sec:jwst}, this discrepancy indicates that our models may underestimate the outward expansion of Fe-rich ejecta, possibly due to an underestimation of the initial explosion asymmetries.

In previous studies, we used model B18.3 to predict the XRISM Resolve spectrum of \sna\ as observed during the allocated performance verification phase in 2024 (\citealt{2024ApJ...961L...9S, 2024RNAAS...8..156S}). This analysis demonstrated that shocked ejecta play already a key role in shaping the emission line profiles, significantly broadening them. 
According to our interpretation, the observed broadening results from the bulk motion of rapidly expanding ejecta along the LoS, contributing to the overall velocity dispersion of the emitting shocked ejecta.

   \begin{figure*}
   \centering
   \includegraphics[width=0.85\textwidth]{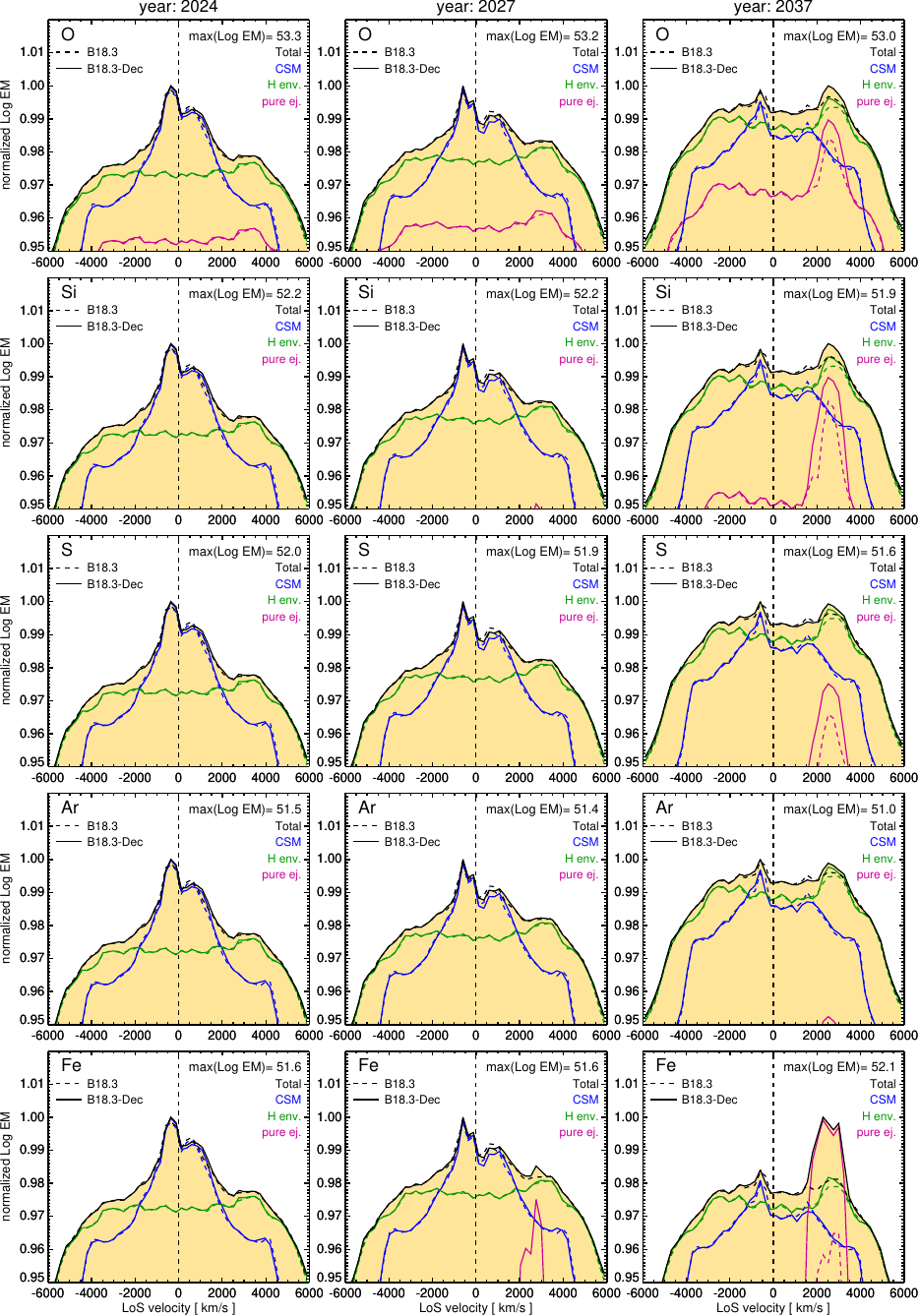}
   \caption{Normalized EM distributions (in log scale, with the maximum value indicated in the upper right corner of each panel; the EM is in units of cm$^{-3}$) as a function of LoS velocity for (from top to bottom) shocked O, Si, S, Ar, and Fe, as derived from models B18.3 (dashed lines) and B18.3-Dec (solid lines) at three epochs: 2024 (left panels), 2027 (center panels), and 2037 (right panels). Each panel displays the total distribution (black line) along with individual contributions from the species in the shocked CSM (blue), the shocked H envelope (assuming CSM-like abundances; green), and the shocked inner (pure) ejecta (violet).}
   \label{em_prof}%
   \end{figure*}

It is therefore insightful to compare the LoS velocity profiles of ejecta species with those of the same elements in the shocked CSM, with the aim to investigate the detectability of shocked ejecta in future observations. For the ejecta, we differentiate between elements originating from the H envelope (assuming similar abundances as the CSM) and those from the metal-rich (in the following "pure") ejecta. Figure~\ref{em_prof} shows the EM distribution as a function of LoS velocity for shocked O, Si, S, Ar, and Fe, as derived from models B18.3 and B18.3-Dec in different epochs. The figure distinguishes between contributions from the CSM, the H envelope, and the pure ejecta, providing a comprehensive view of how these components shape the velocity structure of the remnant. 

Although these EM profiles cannot be directly compared to the emission line profiles observed with high-spectral-resolution instruments such as XRISM Resolve, they provide valuable insights into the expected line broadening due to dynamical effects. In particular, they highlight the influence of shock interaction on the velocity dispersion of different plasma components and help to identify potential spectral features that could be used to diagnose ejecta properties. Furthermore, comparing the contributions of shocked CSM and shocked ejecta enables us to distinguish ejecta-dominated emission in observed line profiles, which is critical for interpreting upcoming high-resolution X-ray spectra (see Sect.~\ref{sec:xrism}). 

As noted by \cite{2024ApJ...961L...9S}, our models indicate that the emission from shocked ejecta (both the H envelope and pure ejecta) is significantly broader than that from the shocked CSM (compare the green/violet lines with the blue lines in Fig.~\ref{em_prof}). Before 2021, the X-ray emission was primarily dominated by the shocked CSM (see Paper I), with line profiles reflecting the distribution of LoS velocities within this component. By 2021, the contribution from shocked ejecta became comparable, and in subsequent years, it continued to grow relative to the shocked CSM, eventually becoming the dominant component (see Paper I).

By 2024 (left panels in Fig.~\ref{em_prof}), at the time of the XRISM observations of \sna\ (OBS ID 300021010), the shocked CSM still dominated the core of the velocity distributions for all species, with velocities ranging between $-2000$ and 2000 km s$^{-1}$. However, the wings of the distributions were increasingly influenced by the contribution from the shocked H envelope, which extended to higher velocities. Similar results were found for both models B18.3 and B18.3-Dec, as the shocked ejecta have not yet been significantly influenced by either the asymmetry of the SN explosion or the effects of the Ni-bubble (compare the solid and dashed lines in the left panels of the figure). The contribution of shocked ejecta is expected to manifest as an additional broadening in both redshifted and blueshifted wings of X-ray emission lines (\citealt{2024ApJ...961L...9S, 2024RNAAS...8..156S}). 

In the coming years, the contribution from shocked ejecta will continue to increase, leading to further broadening of the wings in X-ray emission lines. By 2027 (center panels in Fig.~\ref{em_prof}), the overall distributions remain similar to those of 2024, but with a noticeable rise in the contribution from the shocked H envelope. The most intriguing development is the emergence of a small shocked pure Fe component at velocities between 2000 and 3000 km s$^{-1}$, visible in model B18.3-Dec. This feature results from the interaction of the reverse shock with the Fe-rich clump propagating southward and receding from the observer (see Sect.~\ref{sec:jwst}). In contrast, this feature is less prominent in model B18.3 (see dashed lines in the right panels of Fig.~\ref{em_prof}), as the Fe-rich ejecta remain more compact due to the absence of Ni-bubble effects, which would otherwise enhance their expansion. It is also worth noting that a similar contribution at negative velocities is absent, as the Fe-rich clump propagating northward towards the observer (blueshifted) is less extended and has not yet interacted with the reverse shock in the epoch considered.

By 2037 (right panels in Fig.~\ref{em_prof}), the EM distributions of all species will be almost entirely dominated by shocked ejecta, primarily originating from the H envelope. Additionally, the contribution from shocked pure ejecta will become increasingly significant, particularly in the redshifted wing of the distributions, where a secondary bump will emerge at velocities around $3000$~km~s$^{-1}$. This feature is more pronounced in the distributions of Fe, Si, and S and is stronger in model B18.3-Dec compared to model B18.3 (compares dashed and solid lines in Fig.~\ref{em_prof}). The effect results from the interaction of the reverse shock with the Fe-rich clump, which is mixed with surrounding ejecta and propagates southward, receding from the observer. A similar contribution at negative velocities remains absent also at this epoch, as the Fe-rich clump moving northward toward the observer has not yet significantly interacted with the reverse shock. The contrasting behavior of the two clumps, a direct consequence of the asymmetric bipolar explosion, underscores the lasting influence of the explosion geometry on the remnant’s evolution.

\begin{figure*}
   \centering
   \includegraphics[width=0.95\textwidth]{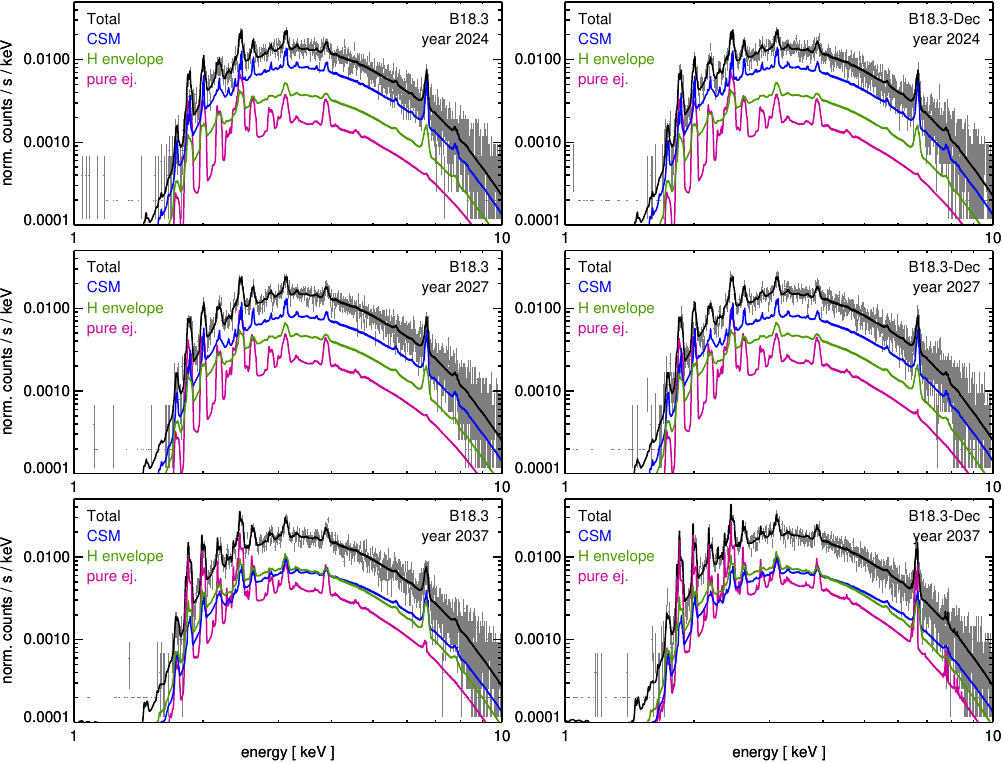}
   \caption{Full XRISM Resolve spectra in the $[1.0, 10.0]$ keV energy range synthesized from models B18.3 (on the left) and B18.3-Dec (on the right) at the labeled times. The spectra (shown as grey crosses) were computed assuming an exposure time of 500~ks. For reference, the figure also shows the ideal synthetic spectra (black lines) along with the contributions from individual plasma components: shocked CSM (blue), shocked H envelope (green), and shocked pure ejecta (violet). No significant signal is detected below 1.8~keV due to the XRISM gate valve remaining closed, which prevents soft X-ray photons from reaching the detector.}
   \label{xrism_spec}%
   \end{figure*}

\subsubsection{Properties of X-ray spectra in the coming years}
\label{sec:xrism}

   \begin{figure*}
   \centering
   \includegraphics[width=0.97\textwidth]{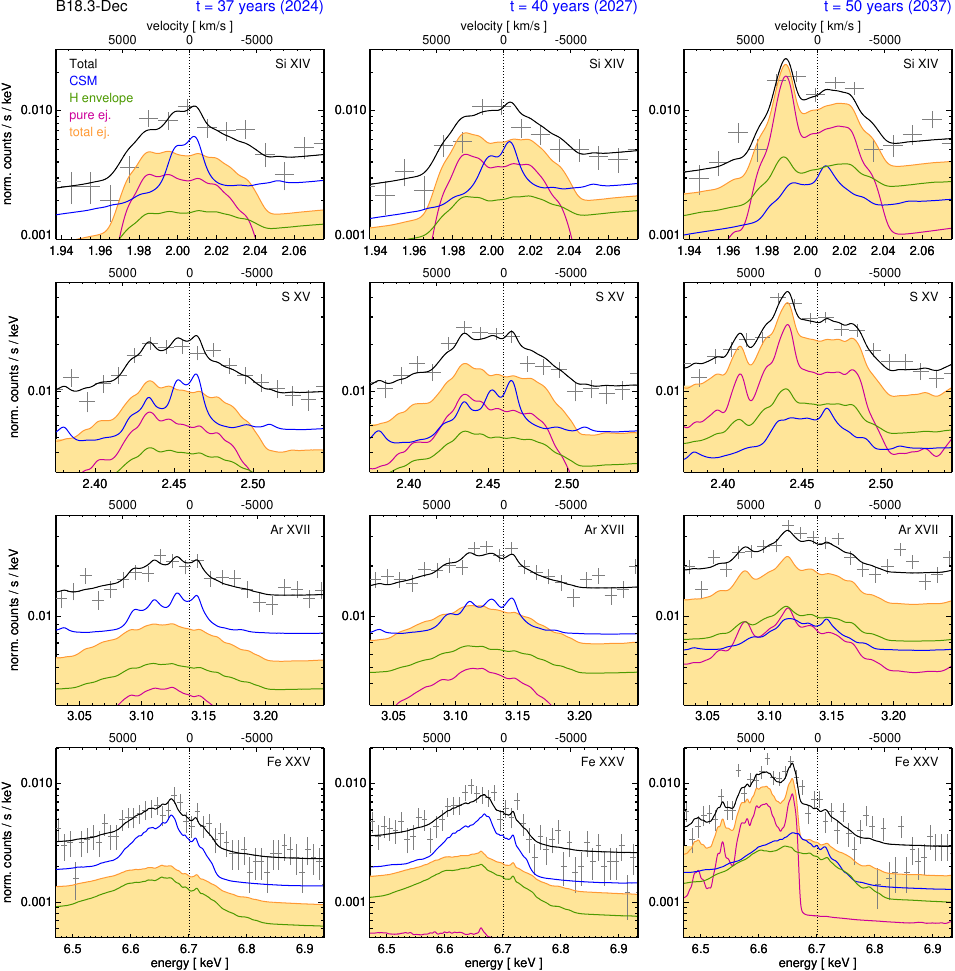}
   \caption{XRISM Resolve spectra synthesized from model B18.3-Dec at three epochs: 2024 (left panels), 2027 (center panels), and 2037 (right panels). The figure presents close-up views of selected emission lines: Si\,XIV (2.1 keV), S\,XV (2.46 keV), Ar\,XVII (3.14 keV), Fe\,XXV (6.7 keV). Each panel shows the synthetic spectra of \sna\ as observed with XRISM Resolve (shown as grey crosses), assuming an exposure time of 500 ks. Also included are the ideal synthetic spectra (black lines) and the contributions from different plasma components: shocked CSM (blue), shocked H envelope (green), and shocked pure ejecta (violet). The total contribution from shocked ejecta (H envelope + pure ejecta) is highlighted by the shaded yellow region. The top axis of each panel indicates the corresponding Doppler shift velocities relative to the rest-frame wavelength of each line.}
   \label{xrism_lines}%
   \end{figure*}

The above findings highlight the need for a detailed assessment of XRISM's diagnostic capabilities in studying the shocked ejecta in \sna. To this end, we synthesized XRISM Resolve spectra from our models at the three epochs analyzed in Fig.~\ref{em_prof}: 2024 (remnant age of 37 years), 2027 (40 years), and 2037 (50 years). For this analysis, we assumed an exposure time of 500~ks for the synthetic spectra, approximately 30\% longer than the $\sim 350$~ks allocated for the performance verification phase in 2024 (OBS ID 300021010). The resulting synthetic spectra, covering the full $[1.0, 10.0]$~keV energy band, are presented in Fig.~\ref{xrism_spec}. Note that emission below 1.8~keV is significantly attenuated due to a technical issue related to XRISM Resolve’s gate valve, which failed to open as originally planned. This anomaly leads to a reduced soft X-ray response, limiting the spectral coverage at lower energies.

Before to proceed, it is important to note that the CSM geometry and density distribution in our models are constrained by X-ray observations collected before 2016 (see Paper I). Consequently, only the shocked CSM as of 2016 is well constrained, while the regions still unshocked at that epoch were extrapolated. This means that if the remnant begins interacting with a CSM that deviates significantly from our extrapolation, substantial differences from our predictions may arise. Such discrepancies would be crucial for refining the models and achieving a more accurate characterization of the CSM.

For our purposes, we focused here on a selection of emission lines that can serve as important diagnostics of the shocked ejecta: Si\,XIV (2.1 keV), S\,XV (2.46 keV), Ar\,XVII (3.14 keV), Fe\,XXV (6.7 keV). Figure~\ref{xrism_lines} shows these lines at the three epochs for model B18.3-Dec; a corresponding figure for model B18.3 would yield nearly identical results, except for differences in the Fe lines (as also suggested by Fig.~\ref{em_prof}). Therefore, our analysis primarily focuses on model B18.3-Dec, highlighting any relevant discrepancies observed in model B18.3.

The figure illustrates the distinct contributions to the emission, particularly from the shocked CSM, the shocked H envelope (assuming CSM-like abundances), and the shocked pure (metal-rich) ejecta, with abundances derived from the simulated evolution of the post-core-collapse ejecta. By examining the evolution of these spectral features across the three epochs, we assessed how XRISM Resolve (and, more broadly, high-resolution X-ray spectroscopy) can differentiate between shocked ejecta and shocked CSM while tracking the progressive enrichment of the X-ray-emitting plasma by metal-rich material. These insights will be valuable for interpreting future high-resolution X-ray observations of \sna\ and other young SNRs.

Figure~\ref{xrism_spec} clearly demonstrates that, in the coming years, the contribution from shocked ejecta will gradually increase. This trend is particularly evident in spectral lines below 3.5~keV, such as Si\,XIV and S\,XV, where the ejecta already make a significant contribution in 2024 and are expected to dominate by 2027 (see the top two rows of Fig.~\ref{xrism_lines}). For these lines, the majority of the ejecta contribution originates from pure ejecta, with the contribution from the H envelope remaining relatively minor (compare violet and green lines in the figure).

At higher energies, the contribution from shocked ejecta is less pronounced but still detectable. This is primarily due to the smaller contribution from pure ejecta compared to low-energy lines (see the bottom two rows of Fig.~\ref{xrism_lines}), as ejecta rich in Ar and Fe are located in the innermost regions of the remnant and interact with the reverse shock at later times. By 2027, the ejecta contribution is expected to become comparable to that of shocked CSM in lines as Ar\,XVII (compare yellow and blue lines in the figure), mainly driven by the shocked H envelope (green line). By 2037, the emission will be largely dominated by shocked pure ejecta. Beyond this point, it is also expected to dominate in high-energy lines (violet line).

As noted by \cite{2024ApJ...961L...9S}, according to our models, the contribution from shocked ejecta can be revealed in spectra collected at the present time by analyzing the line profiles. While the core of the lines remains dominated by the contribution from shocked CSM, the wings show the emerging influence of the broader component from shocked ejecta (see Fig.~\ref{xrism_lines}). This effect is particularly evident in lines such as Si\,XIV and S\,XV, whereas lines at higher energies (e.g., Ar\,XVII and Fe\,XXV) show a less pronounced broadening. The broader wings are a direct consequence of the dynamical properties of shocked ejecta, which exhibit a greater spread of LoS velocities compared to the shocked CSM (see Sect.~\ref{sec:dyn_sh_ejecta} and Fig.~\ref{em_prof}). Notably, the centroids of the lines are generally redshifted, with the redshifted wing extending further than the blueshifted counterpart, consistently with our findings in Sect.~\ref{sec:ejecta_prop}. This effect is most prominent in the Fe\,XXV line, which remains definitively redshifted at all epochs, with velocities ranging between 1000 and 2000~km~s$^{-1}$. This asymmetry mirrors the spatial imbalance of the SN explosion, which released more energy in the direction of the Fe-rich clump receding from the observer (see Sect.~\ref{sec:jwst}).

In the coming years, line broadening is expected to increase as the contribution from shocked ejecta grows, introducing additional asymmetries and velocity shifts in the line profiles. This evolution will enhance the distinct signatures of the ejecta, particularly in the line wings, offering deeper insights into the remnant’s ejecta dynamics and asymmetries. Notably, a prominent redshifted component with a velocity of around 2500 km s$^{-1}$, attributed to pure ejecta, is expected to emerge in the Si XIV and S XV lines, eventually becoming their dominant peak by 2037.

These developments will provide a new means to probe the 3D structure of ejecta and kinematics of the remnant. As the contribution from shocked ejecta increases, elemental abundances derived from spectral analysis will likely become more pronounced, further distinguishing shocked ejecta from shocked CSM. The evolving line profiles will serve as key diagnostics for tracking the dynamics and composition of the shocked material over time. These findings underscore the critical role of high-resolution X-ray spectroscopy in disentangling the contributions of shocked ejecta and CSM, offering a powerful tool to investigate both the explosion physics and the properties of the surrounding medium (see also \citealt{2024arXiv240812462O}).

   \begin{figure}
   \centering
   \includegraphics[width=0.45\textwidth]{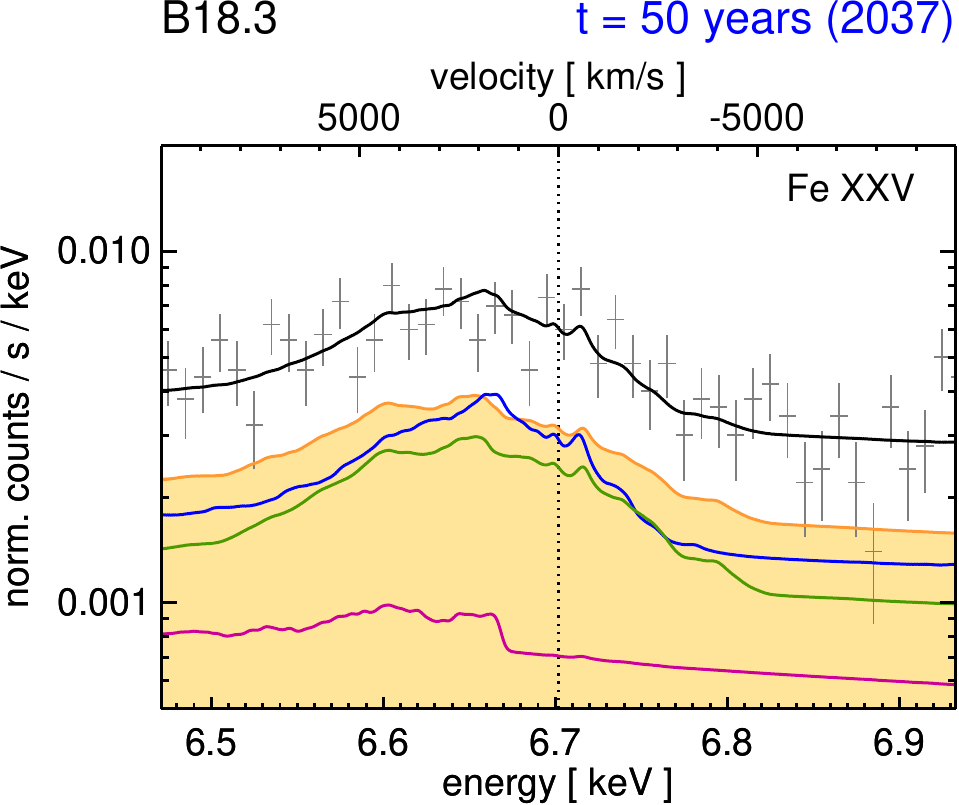}
   \caption{As the bottom right panel in Fig.~\ref{xrism_lines} at the age of 50 years (corresponding to 2037) but for the Fe\,XXV line synthesized from model B18.3.}
   \label{line_Fe_2037}%
   \end{figure}

The effect of the Ni-bubble is not particularly evident in the spectra at the three epochs analyzed, except for Fe lines. In model B18.3-Dec, the contribution from pure Fe ejecta becomes significant between 2027 and 2037 (see lower panels in Fig.~\ref{xrism_lines}), leading to a pronounced redshifted Fe XXV line. In contrast, in model B18.3, the contribution from pure Fe ejecta remains negligible until 2037 (see Fig.~\ref{line_Fe_2037}). This difference arises because, in model B18.3-Dec, the bulk of Fe-rich plume propagating southward, away from the observer, starts to interact with the reverse shock around 2027, whereas in model B18.3, the bulk of Fe-rich clumps is still far from the reverse shock in 2037.

In Sect.~\ref{sec:jwst}, we demonstrated that our model underestimates both the expansion velocity and the extent of the Fe-rich ejecta toward the reverse shock, when compared to JWST observations. As a result, the interaction between the pure Fe-rich ejecta and the reverse shock is delayed by more than 5 years in our models. This suggests that the contribution from this ejecta component, expected to appear between 2027 and 2037, might already be detectable in current data. 
Based on our previous findings, we anticipate that our model underestimates the flux of the Fe XXV line because the Fe-rich ejecta have already begun interacting with the reverse shock, contrary to our model’s predictions. This interpretation is supported by JWST observations, which show Fe-rich material engaging with the reverse shock (\citealt{2023ApJ...949L..27L}). Future models addressing this discrepancy in the distribution of innermost Fe-group elements are unlikely to significantly alter the structure of light and intermediate-mass elements. Therefore, we expect the predicted profiles and fluxes of lines from intermediate-mass elements to remain unchanged.

\subsection{Predicted long-term evolution over the next 5000 years}

We extended our simulations to cover the next 5000 years of evolution, providing insights into the future development of \sna. However, some caution is necessary when interpreting these results. The structure of the CSM beyond the hourglass-shaped H\,II nebula remains largely unknown. As a result, we had to make an arbitrary assumption regarding the extended medium through which the blast wave propagates. To minimize additional uncertainties, we adopted the simplest possible configuration: a surrounding environment with a density profile decreasing as $r^{-2}$ beyond $\approx 2.2\times 10^{18}$~cm (see Sect.~\ref{sec:snrmod}). This choice, which excludes possible inhomegeneities in the CSM and the transition from the CSM to the interstellar medium, prevents the introduction of additional asymmetries in the remnant morphology that could arise from interactions with an arbitrary inhomogeneous medium.

By following this approach, we were able to evaluate how long the remnant retains the imprint of the initial asymmetries introduced by the bipolar explosion, as well as those shaped by its interaction with the highly structured CSM (i.e., the triple-ring nebula around \sna) immediately surrounding the SN. These extended simulations provide a framework for understanding the possible long-term evolution of \sna\ and enables comparisons between its predicted future morphology and that of more evolved SNRs observed today. Identifying features in the morphology of present-day evolved remnants analogous to those predicted for the future of \sna\ could signal early interactions between the blast wave and a highly inhomogeneous CSM similar to that of \sna.

\begin{figure*}
   \centering
   \includegraphics[width=0.9\textwidth]{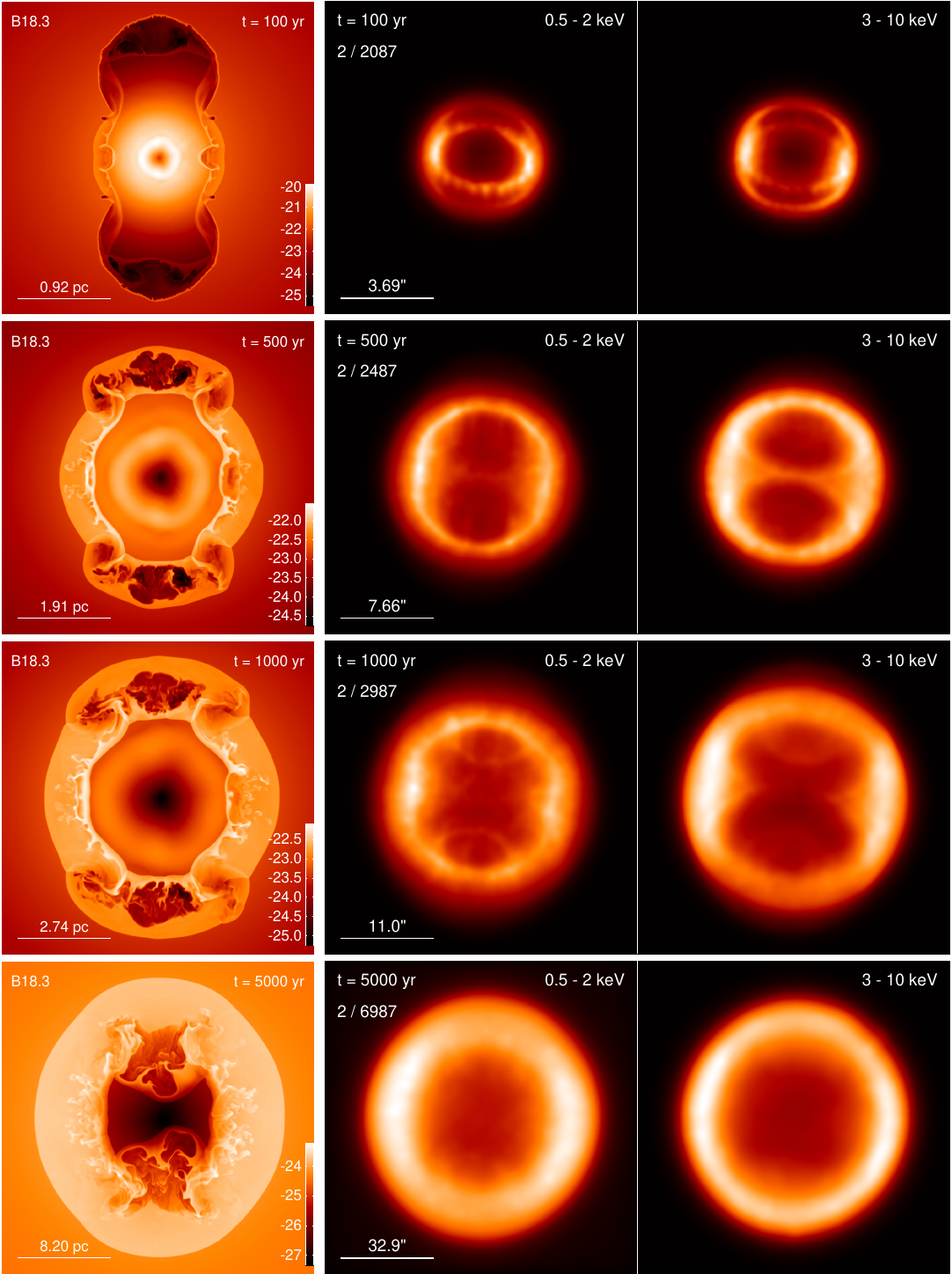}
   \caption{Predicted long-term evolution of \sna\ over the next 5000 years based on model B18.3. The left panels show cross-sections of the mass density distribution (in log scale, in units of g~cm$^{-3}$) in a plane perpendicular to the equatorial ring, passing through the explosion center, at four different epochs (see labeled times). The right panels present the corresponding synthetic thermal X-ray emission maps in the $[0.5, 2.0]$~keV (on the left) and $[3.0, 10.0]$~keV (on the right) bands, assuming an orientation consistent with that of \sna. Each X-ray image is normalized to its maximum for visibility, and the maps are convolved with a Gaussian of 0.1 arcsec to approximate the spatial resolution of Chandra.}
   \label{future_sn87a}%
   \end{figure*}

\begin{figure}
   \centering
   \includegraphics[width=0.48\textwidth]{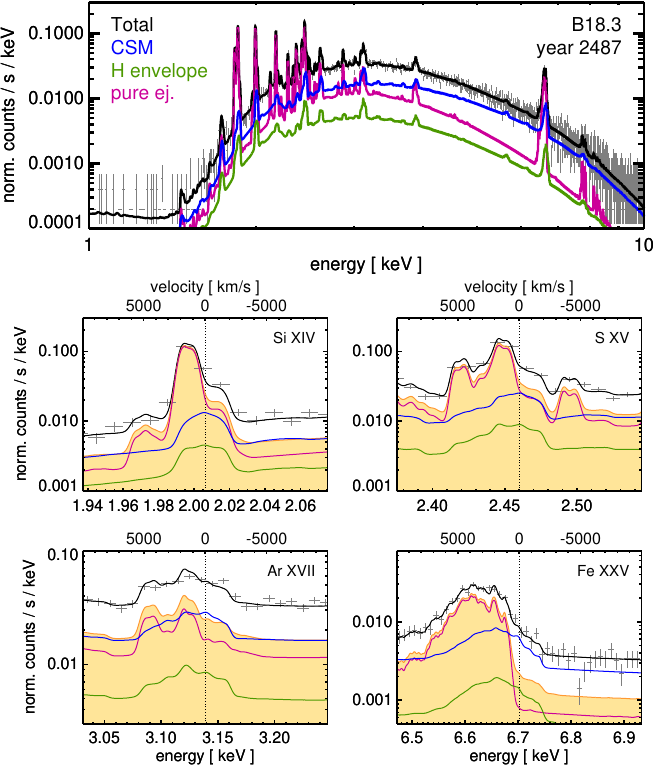}
   \caption{XRISM Resolve spectra synthesized from model B18.3 at an age of 500 years (corresponding to the year 2487). The upper panel displays the full spectrum in the $[1.0, 10.0]$~keV energy range, while the four lower panels provide close-up views of selected emission lines: Si\,XIV (2.1 keV), S\,XV (2.46 keV), Ar\,XVII (3.14 keV), Fe\,XXV (6.7 keV). The top axis of each lower panel represents the corresponding Doppler shift velocities relative to the rest-frame wavelength of each line. The figure includes the ideal synthetic spectra (black lines) and the contributions from individual plasma components: shocked CSM (blue), shocked H envelope (green), and shocked pure ejecta (violet). The lower panels emphasize the contribution from shocked ejecta (H envelope + pure ejecta) using a shaded yellow region.}
   \label{xrism_500years}%
   \end{figure}

Figure~\ref{future_sn87a} illustrates the evolution of \sna\ at four selected epochs, spanning from 100 to 5000 years after the explosion, as predicted by model B18.3. The results for model B18.3-Dec are qualitatively similar, so we primarily focused on model B18.3. The left panels in the figure display 2D slices of the remnant, taken in a plane perpendicular to the equatorial ring and passing through the explosion center. These slices provide insight into the internal structure of the remnant, including the evolution of the shocked ejecta and the mixing region between the forward and reverse shocks. The center and right panels show the corresponding synthetic X-ray images in the $[0.5, 2]$ and $[3, 10]$~keV bands, modeled as they would appear in Chandra observations, assuming an orientation consistent with that of \sna. These images highlight the changing morphology of the remnant over time and allow for direct comparison with present X-ray morphology of \sna\ and other more evolved SNRs.

At an age of 100 years, the remnant still retains the imprint of its interaction with the dense equatorial ring and the surrounding H\,II region (see upper right panel in Fig.~\ref{future_sn87a}). At this stage, the two outer rings above and below the equatorial plane are partially shocked (see the 2D slice in the upper left panel of Fig.~\ref{future_sn87a}). The remnant’s X-ray morphology remains similar to its present-day appearance, dominated by a bright ring-like structure resulting from the shock interaction with the dense equatorial material. In contrast, emission from the shocked outer rings remains too faint to be detected in X-rays. As the remnant continues to evolve, the contribution of the outer rings to the overall X-ray emission will remain marginal.

As the remnant expands into a progressively lower-density CSM, its morphology undergoes a gradual transformation, shifting from the highly asymmetric structure observed at 100 years (upper panels in the figure) to a more spherically symmetric shape predicted at the age of 5000 years (lower panels). This transition occurs as the expansion reduces density contrasts between different regions and variations in shock velocity average out over larger scales. This process is particularly effective in a CSM with a density profile decreasing as $r^{-2}$, as assumed here, where differences in shock propagation speed naturally diminish over time, further contributing to the remnant's increasing symmetry (see also \citealt{2021A&A...654A.167U}). 

Meanwhile, the reverse shock steadily propagates inward, progressively shocking intermediate-mass and Fe-group elements. By the age of 500 years, Fe-rich regions become prominent as the reverse shock interacts with the two Fe-rich plumes. As an example, Fig.~\ref{xrism_500years} presents the synthetic XRISM Resolve spectrum derived from model B18.3 at an age of 500 years, assuming an exposure time of 500~ks. At this epoch, the contribution of shocked ejecta is significant across all emission lines and becomes the dominant component in softer-energy lines (e.g., Si XVI and S XV). Notably, a prominent Fe K$\alpha$ line at 6.7 keV appears, significantly redshifted, and originating from the shocked ejecta. This feature emerges despite the fact that the model considered (B18.3) does not account for Ni-bubble effects. The Fe line is primarily due to the interaction of a substantial fraction of the Fe-rich clumps with the reverse shock at this epoch.

The imprint of the remnant's early interaction with the dense equatorial ring and the surrounding H\,II region remains visible in X-ray images for up to approximately 1000 years (see second and third rows in Fig.~\ref{future_sn87a}). The central bar-shaped structure, most prominent at 500 years (particularly in the $[3, 10]$~keV band), originates from this early interaction with the dense equatorial material. Although it becomes less pronounced over time, this feature remains partially visible at 1000 years, highlighting the lasting influence of the asymmetric CSM on the remnant’s morphology.

Similar structures in older SNRs offer valuable clues about their early evolution. Such features may indicate interactions between the SN shock and a circumbinary disk, similar to what is observed in \sna, emphasizing the role of pre-SN mass loss in shaping SNR structures. Notable examples include: G292.0+1.8, which exhibits a central bar-like structure, likely shaped by circumstellar interactions; E0102-72.3 (1E 0102.2-7219), whose morphology suggests the progenitor exploded within a disk of material, denser in the equatorial plane; SNR 0540-69.3, characterized by an outer shell and bright central structures, likely influenced by interactions with surrounding material.

By 5000 years, the remnant’s X-ray morphology is predicted to become nearly spherically symmetric, with no discernible traces of its early interaction with the equatorial ring, marking the transition to a more uniform expansion phase.

\section{Summary and Conclusions}
\label{sec:summary}

In this study, we analyzed high-resolution 3D MHD simulations to investigate the evolution of \sna’s ejecta structure, focusing on explosion asymmetries and ejecta-CSM interactions. By comparing our models with recent JWST observations and making predictions for XRISM data, we examined how the initial asymmetric explosion shaped the observed ejecta structure, identifying constraints for future modeling. We also explored the impact of Ni-bubble effects on Fe-rich ejecta expansion and predicted future X-ray signatures for diagnostic purposes. Additionally, we extended our simulations to 5000 years to determine how long the remnant retains imprints of the explosion and its early interaction with the inhomogeneous CSM. Our key findings are summarized below.

\begin{itemize}
\item {\bf Fe-rich plumes and bipolar explosion.} The simulations successfully reproduced the large-scale Fe-rich ejecta morphology, revealing the formation of two prominent Fe-rich plumes or clumps. According to our models, these clumps originate from an asymmetric bipolar explosion, which is able to reproduce Fe line profiles observed 1\textendash 2 years after the SN (e.g., \citealt{1990ApJ...360..257H}). Their motion is consistent with JWST observations, where one clump is moving to the north toward the observer (blueshifted) and the other, more massive clump is moving to the south away from the observer (redshifted).

\item {\bf Impact of Ni-bubble on ejecta evolution.} The Ni-bubble effect significantly influences the dynamics of Fe-rich ejecta, enhancing their expansion and accelerating their interaction with the reverse shock. In the model incorporating the Ni-bubble effect (B18.3-Dec), the interaction of shocked Fe with the reverse shock occurs approximately three years earlier than in models without this effect (B18.3). Furthermore, the Ni-bubble contributes to a broader, smoother LoS velocity distribution of Fe, which peaks at redshifted velocities and partially merges the two Fe-rich clumps into a single, more unified structure.

\item {\bf X-ray diagnostics and future observations.} Our models predict that shocked ejecta have contributed increasingly to X-ray emission since 2021 and will soon dominate as emission from the shocked CSM fades. According to these results, XRISM spectra in 2024 can reveal broad Fe, Si, and S line profiles characteristic of shocked ejecta interactions. We anticipate that future XRISM observations will track the expansion and evolution of these structures, refining constraints on explosion geometry and mixing. These diagnostics will provide key insights into explosion asymmetries, mixing processes, and remnant dynamics, highlighting the crucial role of X-ray spectroscopy in probing both the parent SN and its surrounding environment.

\item {\bf Long-term evolution of the remnant.} Over the next 5000 years, the remnant is expected to evolve from its current asymmetric structure, shaped by interactions with the equatorial ring and H\,II region, to a nearly spherically symmetric morphology, provided it does not encounter significant CSM asymmetries. At 100 years, the remnant will still exhibit a bright, ring-like X-ray structure, due to the early shock interaction with the dense triple-ring nebula. As expansion continues into a lower-density medium (arbitrarily modeled with a simple $r^{-2}$ profile), density contrasts will diminish, promoting symmetry. By 500 years, the emission of a central bar-like feature (the relic of the early interaction with the dense ring) will peak before fading by 1000 years. Finally, by 5000 years, all memory of the early explosion asymmetries and CSM interaction will be lost, leaving behind a nearly homogeneous remnant structure. Similar features may be observed in older SNRs, providing clues about possible early interactions with a structured CSM or circumbinary disk, similar to what is observed in \sna.
\end{itemize}

\subsection*{Tension between current models and JWST observations}

The overall good agreement between our simulations and observations provides compelling evidence that the large-scale structure of \sna's ejecta was shaped by a highly asymmetric core-collapse SN. However, while some aspects are captured by the adopted model \citep[see also][]{2020ApJ...888..111O}, others suggest the need for more extreme or different ejecta properties. Discrepancies that remain include: the underprediction of Fe-rich ejecta velocities of the Fe-rich ejecta by about 35\% compared to those inferred from JWST data, failure to reproduce the observed broken-dipole morphology, and the absence of clearly separated Fe-rich clumps evident in the observations. These tensions suggest that additional physical factors or conditions may need to be incorporated to fully reconcile models with observations.

The question of the too slow ejecta could be related to more extreme explosion asymmetries than considered in our models, which could stem from large-scale instabilities present in neutrino-driven SNe. These include processes such as convective overturn driven by neutrino heating and the SASI in the first seconds after core collapse \citep[e.g.,][]{2003ApJ...584..971B, 2016ARNPS..66..341J, 2016PASA...33...48M, 2020LRCA....6....3M, 2025arXiv250214836J}. These processes are inherently stochastic and can seed diverse and highly asymmetric structures in the expanding ejecta. In addition, the subsequent evolution of the extended Ni-rich plumes is influenced by progenitor properties, which may enhance the acceleration of clumps via Rayleigh-Taylor instabilities, and by interactions with the reverse shock and other reflected shocks that can re-energize the ejecta \citep{2021MNRAS.502.3264G}. Increasing the explosion energy could in principle lead to higher velocities; however, our models already employ an energy of $\sim 2\times 10^{51}$~erg, which is at the high end of values inferred for \sna\ \citep[e.g.,][]{2015ApJ...806..275P}. Furthermore, our B18.3-Dec model accounts for the acceleration induced by radioactive $\beta$-decay heating, which is expected to provide an upper limit to this effect. Nonetheless, all these factors may collectively contribute to higher ejecta velocities and warrant further investigation through dedicated simulations.

The broken-dipole morphology may naturally arise from the intrinsic asymmetries in neutrino-driven explosions. Although multipolar asymmetries are typical, simulations sometimes result in unipolar or approximately bipolar morphologies \citep[e.g.,][]{2015A&A...577A..48W, 2021ApJ...914....4U, 2024ApJ...974...39W, 2024ApJ...964L..16B}, which may offer a potential explanation for the Fe-rich clumps observed by JWST. However, it remains to be demonstrated whether such structures can accurately reproduce the specific distribution of clumps and expansion velocities observed in \sna.

As an alternative to neutrino-driven models, magneto-rotational SNe (MR-SNe) may help to explain some of the observed features better (\citealt{2002ApJ...568..807W, 2003ApJ...584..954A, 2004ApJ...616.1086T, 2009ApJ...691.1360T, 2006A&A...450.1107O, 2009A&A...498..241O, 2014ApJ...785L..29M}). These core-collapse explosions are driven by the rapid rotation and strong magnetic fields of their progenitor stars. In MR-SNe, differential rotation amplifies magnetic fields through the magnetorotational instability, generating powerful magnetic stresses that launch highly energetic and often asymmetric Ni-rich jet-like structures (e.g., \citealt{2021MNRAS.503.4942O}). While MR-SNe can produce hypernova-like energies, potentially contrasting with those observed in \sna, their distinct bipolar ejecta morphologies could provide a plausible explanation for the fast Fe-rich ejecta and the apparent separation of clumps observed by JWST. However, MR-SNe tend to produce well-aligned, collimated bipolar structures. The observed broken-dipole morphology of \sna, consisting of two main clumps separated by approximately $135^\circ$ and an additional smaller clump interacting with the reverse shock \citep{2023ApJ...949L..27L}, appears more complex than standard MR-SN predictions. Further simulations are required to investigate whether MR-SNe can accommodate such deviations from symmetry.

A further challenge for neutrino-driven models is the existence of two clearly distinct Fe-rich clumps in the JWST data (\citealt{2023ApJ...949L..27L}). Whether this apparent separation reflects an intrinsic feature of the ejecta or arises from observational limitations (e.g., limited sensitivity to the remnant's central regions) remains an open question. In general, neutrino-driven simulations predict centrally peaked Fe distributions, without persistent large-scale separations. Even the B18.3-Dec model presented here, which initially shows two clumps due to asymmetric ejection, ultimately yields a merged structure after expansion and decay heating. In contrast, early ($t<10$ s) MR-SN simulations \citep[e.g.,][]{2021MNRAS.503.4942O, 2023MNRAS.518.1557R} reveal highly asymmetric ejecta, with Fe-rich components concentrated in apparently well-separated lobes, which may account for the two Fe-rich clumps observed by JWST. However, the long-term evolution of these structures beyond $t\sim10$~s remains largely unexplored and it is unclear whether the observed clumpiness can persist on the age of \sna. 

Taken together, these considerations suggest that the Fe distribution in the central regions of the remnant may provide a key diagnostic for discriminating between competing explosion scenarios. Neutrino-driven explosions typically leave behind a centrally concentrated Fe-rich component, while MR-SNe may produce reduced central Fe abundance due to efficient evacuation by jets. Thus, future high-resolution observations of the remnant’s center could be decisive in discriminating between these scenarios for \sna.

Despite their potential to explain several features of \sna, MR-SNe also present substantial challenges. First, the progenitor of \sna\ (a BSG likely resulting from binary evolution; \citealt{2007Sci...315.1103M}) does not satisfy the typical conditions associated with MR-SNe, which usually require rapidly rotating progenitors with strong pre-collapse magnetic fields \citep[e.g.,][]{2010ApJ...719L.204W, 2015Natur.528..376M, 2021MNRAS.500.4365A}. Second, \sna\ behaves like a standard Type IIP SN in terms of its explosion properties, with the unusual light curve originating from the loss of H during the progenitor's merger (\citealt{2021ApJ...914....4U}). Third, the observed ejecta morphology, while suggestive of bipolarity, does not conform to the highly collimated jets expected from MR-SNe. Finally, MR-SNe are thought to produce highly magnetized compact remnants (magnetars; e.g., \citealt{2014ApJ...785L..29M, 2021MNRAS.500.4365A}). Although recent studies suggest the presence of a compact object near the remnant's center \citep[e.g.,][]{2019ApJ...886...51C, 2021ApJ...908L..45G, 2022ApJ...931..132G, 2024Sci...383..898F}, the observed emission is less energetic than expected from magnetars, which typically exhibit high spin-down energy outputs (\citealt{2017ARA&A..55..261K, 2025arXiv250304442R}). Such a source would either be too bright at present or would have already spun down, depositing its energy earlier and making \sna\ significantly brighter than observed. 

In summary, while both neutrino-driven and magnetorotational explosion models offer viable pathways to account for several observed features of \sna, neither scenario fully captures the complexity revealed by current observations. This tension underscores the necessity for more advanced simulations and complementary observational efforts. High-resolution, multi-dimensional models exploring a broader range of explosion geometries, including more extreme asymmetries (Ono et al., in prep.), will be key to resolving the discrepancies between existing models and the JWST data. Unraveling the origin of the nearly bipolar morphology of \sna\ will critically depend on these developments.

In the coming years, the synergy between next-generation instruments, including JWST and XRISM (along with future missions like AXIS and NewAthena), and increasingly sophisticated theoretical models will offer unprecedented constraints on the kinematics, morphology, and composition of the ejecta in \sna. In particular, monitoring the evolution of Fe-rich clumps as they interact with the reverse shock, characterizing the X-ray emission from shocked ejecta, and investigating potential non-thermal emission from a pulsar wind nebula will be crucial diagnostics. These efforts will not only shed light on the origin of the explosion asymmetries in \sna\ but also provide valuable insights into the physical mechanisms driving core-collapse SNe more broadly.

\begin{acknowledgements}
We thank an anonymous referee for the useful suggestions that allowed us to improve the manuscript. S.O. is grateful to Thomas Janka (Max-Planck-Institut f\"ur Astrophysik) for insightful discussions on the physics of neutrino-driven supernovae and the evolution of \sna. The \PLUTO\ code is developed at the Turin Astronomical Observatory (Italy) in collaboration with the Department of General Physics of  Turin University (Italy) and the SCAI Department of CINECA (Italy). We acknowledge that part of the results of this research have been achieved using the PRACE Research Infrastructure resource Marconi based in Italy at CINECA (PRACE Award N.2016153460). Additional computations were carried out on the HPC system Leonardo hosted at CINECA (HP10BUMIQR) and on the HPC system MEUSA at the SCAN (Sistema di Calcolo per l'Astrofisica Numerica) facility for HPC at INAF-Osservatorio Astronomico di Palermo. S.O., M.M., F.B., and V.S. acknowledge financial contribution from the PRIN 2022 (20224MNC5A) - ``Life, death and after-death of massive stars'' funded by European Union – Next Generation EU. S.O., M.M., F.B., E.G., and O.P. acknowledge financial contribution from the INAF Theory Grant ``Supernova remnants as probes for the structure and mass-loss history of the progenitor systems''. 
M.M. acknowledges support by the  Fondazione ICSC, Spoke 3 Astrophysics and Cosmos Observations. National Recovery and Resilience Plan (Piano Nazionale di Ripresa e Resilienza, PNRR) Project ID CN\_00000013 "Italian Research Center on  High-Performance Computing, Big Data and Quantum Computing" funded by MUR Missione 4 Componente 2 Investimento 1.4: Potenziamento strutture di ricerca e creazione di "campioni nazionali di R\&S (M4C2-19 )" - Next Generation EU (NGEU).
S.N. is supported by the JST ASPIRE project for top scientists, ``RIKEN-Berkeley Mathematical Quantum Science Initiative''.
SHL is supported by JSPS grant No. JP24K07092 and the World Premier International Research Center Initiative (WPI), MEXT, Japan.
O.P. acknowledges partial support through the MSCA4Ukraine project from the European Union. Views and opinions expressed are however those of the authors only and do not necessarily reflect those of the European Union. Neither the European Union nor the MSCA4Ukraine Consortium as a whole nor any individual member institutions of the MSCA4Ukraine Consortium can be held responsible for them.
M.-A.A., M.G., G.LM, M.Obe. and B.G acknowledge the support through the grant PID2021-127495NB-I00 funded by MCIN/AEI/10.13039/501100011033 and by the European Union, 
and the Astrophysics and High Energy Physics programme of the Generalitat Valenciana ASFAE/2022/026 funded by MCIN and the European Union NextGenerationEU (PRTR-C17.I1).
M.-A.A. and M.Ono further acknowledge the support through the Generalitat Valenciana via Prometeo excellence programme grant CIPROM/2022/13.
M.G., G.LM, and B.G further acknowledge the support through the Generalitat Valenciana via the grant CIDEGENT/2019/031.

\end{acknowledgements}

\bibliographystyle{aa}
\bibliography{references}

\end{document}